# Dual frequency parametric excitation of a nonlinear multi degree of freedom amplifier with a digitally modified topology

A. Dolev[a,*] and I. Bucher[a]

[a] *Dynamics Laboratory, Faculty of Mechanical Engineering, Technion, Haifa, 3200003, Israel*

Keywords: parametric amplifier, multi degree of freedom, modes of vibration, model updating, tunable structure.

## Abstract

Mechanical or electromechanical amplifiers can exploit the high-Q and low noise features of mechanical resonance, in particular when parametric excitation is employed. Multi-frequency parametric excitation introduces tunability and is able to project weak input signals on a selected resonance. The present paper addresses multi degree of freedom mechanical amplifiers or resonators whose analysis and features require treatment of the spatial as well as temporal behavior. In some cases, virtual electronic coupling can alter the given topology of the resonator to better amplify specific inputs. An analytical development is followed by a numerical and experimental sensitivity and performance verifications, illustrating the advantages and disadvantages of such topologies.

## 1 Introduction

Amplifiers enhance weak physical signals and increase their observability in various fields of engineering [1–4]. Their physical principle by which amplification is achieved differs in accordance with their utilization and role [5], and low loss mechanical amplifiers can perform better than their electronic counterparts [6]. Therefore, the present work focuses on mechanical amplifiers with multiple inputs and outputs, which can handle multiple signals simultaneously. These can be used for multiple antenna transmitters [7] and multiple input multiple output communications [8]. When different inputs act on the multi degree of freedom (MDOF) amplifier at different locations, their response is projected differently on its normal modes [9], allowing for their detection and amplification. Moreover, the MDOF amplifier normal modes can be designed to increase the amplifier gain and sensitivity with respect to specific inputs according to their frequency and point of action.

It has been shown that it is advantageous to utilize parametric excitation to achieve large amplification [10,11]. Additionally, certain parametric excitations, e.g., degenerate amplifiers [12], have a fixed and narrow bandwidth, which allows only amplifying specific frequencies. This property may hinder the ability to amplify a general input signal. For the degenerate amplifier, a signal with a frequency different from twice the natural frequency is not amplified. To overcome this problem, several methods are devised and used.

---

[*] Corresponding author. Tel.: +972-0524-869279.
*E-mail address:* amitdtechnion@gmail.com.

In previous work [13] the dual frequency parametric amplifier (DFPA) scheme was introduced. The scheme utilizes the advantages of two operating modes of parametric amplifiers, a degenerate, and a non-degenerate mode. According to the DFPA scheme, the amplifier is parametrically excited, aka pumped, simultaneously at two algebraically related frequencies $\omega_a$ and $\omega_b$. The degenerate mode is realized by pumping the amplifier at a frequency close to twice the natural frequency $(\omega_n)$, $\omega_a \approx 2\omega_n$, while the non-degenerate mode is realized by pumping it at $\omega_b \approx \omega_n - \omega_r$, where $\omega_r$ is the input signal frequency. While the degenerate mode produces large amplification, it lacks the ability to practically amplify signals with frequencies different from twice one of the amplifier's natural frequencies. The non-degenerate mode, on the other hand, is utilized to amplify monochromatic signals out of a possible wide-band, while producing relatively small amplification. The DFPA benefits from both operating modes, therefore allows to considerably amplify monochromatic signals out of a possibly wide-band, while retaining sensitivity to the input magnitude and phase. This amplifier can be used to detect the effect of unbalance in rotating structures without spinning them at critical speeds [14].

Another method to extend the narrow bandwidth limitation of a typical single degree of freedom degenerated amplifier, is to revise its topology thus increasing the number of degree of freedom. By doing so, the amplifier now has several natural frequencies, therefore several monochromatic inputs can be amplified. However, its topology should be carefully designed to avoid undesired effects such as primary and internal resonances [11,15]. Additionally, care should be taken as to where on the structure the inputs, or forces [5], act and what is measured.

To allow more flexibility, digital, real-time topology modification can be implemented in the amplifier by employing sensors, actuators and a fast-digital signal processor. The incorporation of actuators in a closed-loop allows to modify the stiffness and in fact the topology, leading to some control over the natural frequencies and normal modes. Natural frequency modification is beneficial, because it allows to widen the frequency band of allowed input signals that can be amplified. For example, when operating in the degenerate mode, a certain change in the natural frequency doubles the frequency of the input signal that can be amplified.

Normal modes modification is shown herein to be essential in some cases because it allows to increase sensitivity and observability of certain input patterns. Consider the case of a symmetric system with three DOF, as shown in Figure 1, where the input force acts between $m_1$ and $m_3$, and the sensing is done by measuring the relative displacement between $m_1$ and $m_2$. If the input resonates the system at the third natural frequency, hence the third mode, the input cannot be observed. Therefore, by tuning the stiffness, the second and third modes are exchanged, and the signal can be observed and amplified.

Previous publications [13,14,16] have addressed single degree of freedom systems, or a method affecting a single mode [14]. The present paper expands this idea into MDOF vibrating systems, where all the natural frequencies and normal modes are considered. It has been observed, as reported here, that this expansion requires attention to additional details as some difficulties arise. To avoid some of the difficulties such as internal resonances, a nonlinear optimization procedure was integrated during the mechanical design. A novel approach to circumvent problems such as observability and sensitivity were dealt by incorporating real-time topology modification.

The degenerated operating mode of a lightly damped parametric amplifier is narrow banded in comparison to primary resonance, which is already narrow banded. Therefore, parametric amplifiers can be very sensitive to small modeling errors that influence the estimated natural frequencies. It is demonstrated in the paper that a model updating stage [17], based on measured data improves the accuracy and hence the performance greatly. The model updating approach uses multi-input model identification, a linear model update stage followed by a nonlinear optimization stage.

The paper begins with introduction of the experimental system and description of the problem. Afterwards, the governing equations of motion are developed and solved for the multiple input, multiple output case. The third part deals with numerical validation and experimental verification. Lastly, conclusions and possible implementations are discussed.

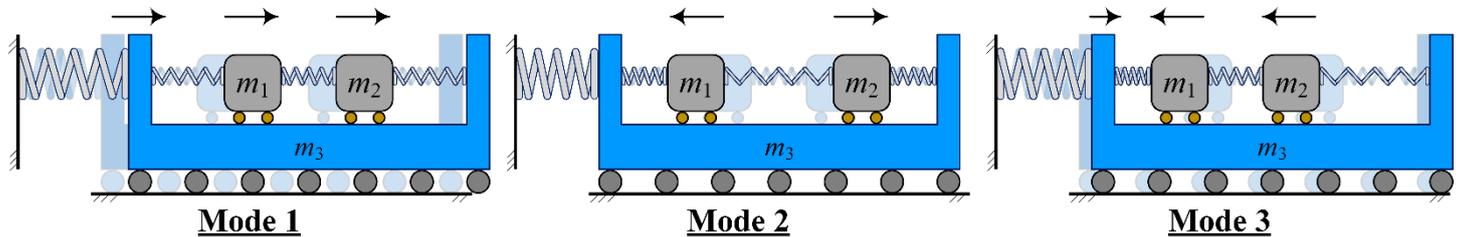

Figure 1. The experimental system mode shapes.

## 2  The experimental rig and setup

In the following, a nonlinear multi degree of freedom system, whose model is shown in Figure 2, is studied. Nonlinearity is caused by a cubic spring added to the system electronically. The system is subjected to several inputs, one at a time, while being pumped (i.e., driven by parametric excitation) with a dual frequency signal according to previous work [13,16]. Additionally, its stiffness and damping are tunable in real-time via a closed-loop signal processor. The ability to sense and amplify input signals is investigated using the methods briefly described in the introduction and in [13,16]. An important attribute of amplifiers, which can be lost when the DFPA scheme is used, is their sensitivity to the input signal phase and amplitude. These sensitivities are very important for some applications (e.g., [14]), therefore, the ability to retain them for different inputs is also studied.

To study the nonlinear MDOF DFPA performances an experimental rig was designed and built according to the model. During the design, a nonlinear optimization procedure was used to corroborate the ability to produce large amplitudes, as described in Section 4, and avoid internal resonances. The rig is depicted in Figure 3, and comprises two modular masses (brass discs can be added or removed) which are connected to a large aluminum plate via leaf springs. The large plate is suspended by additional four leaf springs whose length is adjustable. The modular masses and adjustable springs allow for some structural modifications and adjustments. Two linear voice coil actuators apply forces, and three laser displacement sensors (two Keyence$^{TM}$ LK-H027 and one LS-7030, Osaka, Osaka Prefecture, Japan) measure the displacements (see Figure 4).

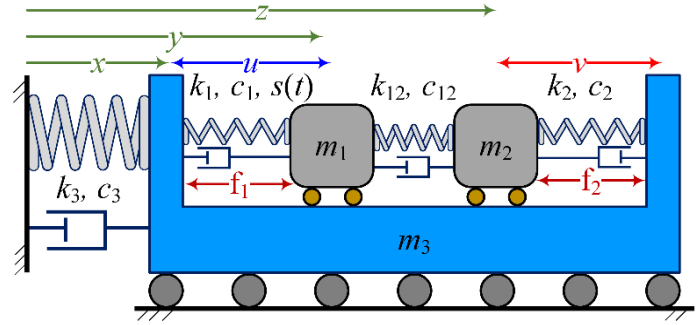

Figure 2. The MDOF amplifier model.

The nonlinear stiffness and stiffness modulation are realized by applying position related forces via the appropriate actuator, while the positions are measured in real-time by the displacement sensors. To easily tune the nonlinear stiffness and stiffness modulation a real-time digital signal processor (DSP; dSPACE 1104) is used. The DSP serves as a data acquisition device and as a displacement dependent function generator. A flow-chart of the experimental setup including the peripheral devices is shown in Figure 4.

The topology modification presented in this paper, includes the addition and subtraction of a coupling link between $m_1$ and $m_2$, which comprises a spring $(k_{12})$ and a dashpot $(c_{12})$, as shown in Figure 2. This link was added via the actuators, by applying forces which are proportional to the masses' relative position and velocity.

Current amplifiers are incorporated into the actuators' circuits to simplify the model and the forces determination scheme, because the forces applied by a linear voice coil are (nominally) linearly proportional to the induced current [18].

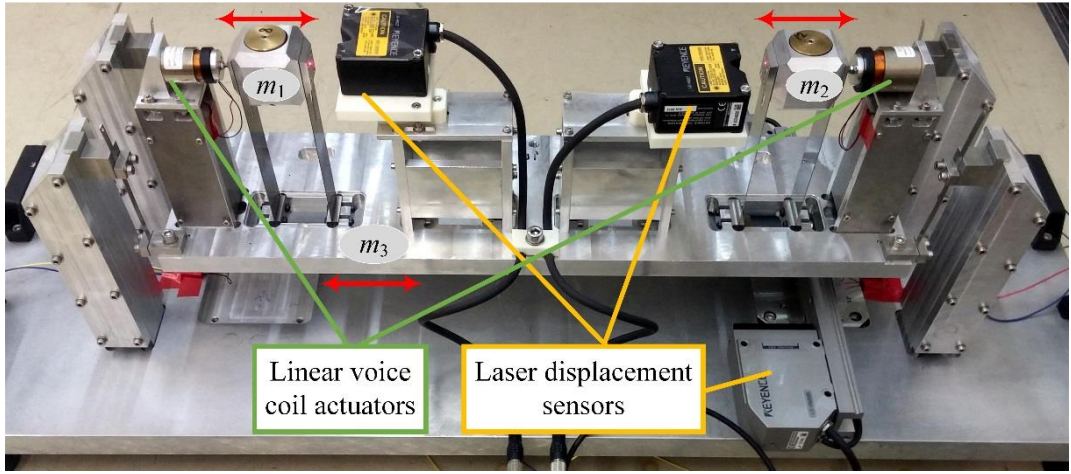

Figure 3. The Experimental rig.

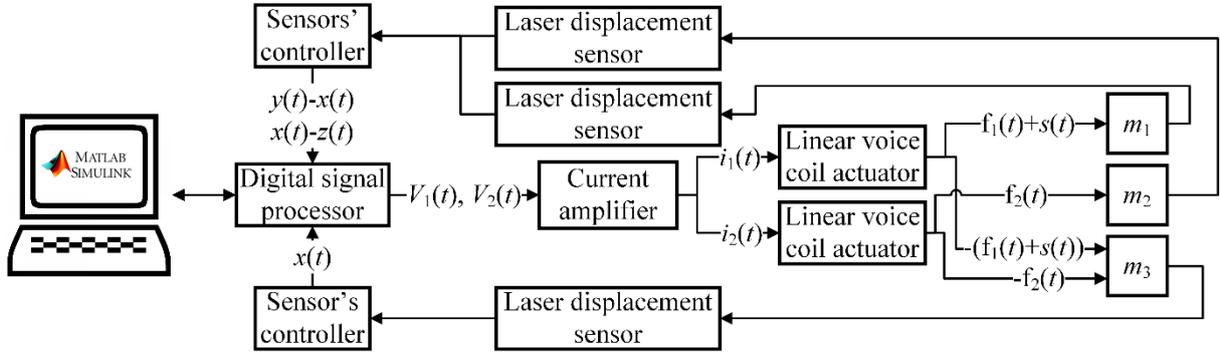

Figure 4. Experimental rig setup flow-chart.

## 3 Mathematical derivation of the governing Equations of motion

The model of the MDOF dual frequency parametric amplifier is presented in this section alongside the governing equations of motion (EOM). The dimensional governing EOM are first derived using Hamilton's principle. Then, these EOM are transformed to scaled EOM in modal coordinates using several transformations and definition of new parameters.

### 3.1 Governing EOM in physical coordinates

The governing EOM of a tunable, nonlinear parametrically excited amplifier having three DOF are derived in this section. The studied DFPA shown in Figure 3 is a representative case of a MDOF parametric amplifier, and its model is depicted in Figure 2. The DFPA is characterized by particle masses $m_\bullet$, linear dashpots $c_\bullet$, linear stiffness $k_\bullet$, and time varying stiffness $s(t)$. In addition, two external forces can be applied between the masses $m_1$ and $m_3$ and between $m_2$ and $m_3$, which are $f_1$ and $f_2$ accordingly. It proves convenient to define relative coordinates (see Figure 2):

$$u = y - x, \quad v = x - z. \tag{1}$$

The derived EOM using Hamilton's principle and the relative coordinates are as follows:

$$\begin{pmatrix} \sum m_i & m_1 & -m_2 \\ m_1 & m_1 & 0 \\ -m_2 & 0 & m_2 \end{pmatrix} \begin{Bmatrix} \ddot{x} \\ \ddot{u} \\ \ddot{v} \end{Bmatrix} + \begin{pmatrix} c_3 & 0 & 0 \\ 0 & c_1 + c_{12} & c_{12} \\ 0 & c_{12} & c_2 + c_{12} \end{pmatrix} \begin{Bmatrix} \dot{x} \\ \dot{u} \\ \dot{v} \end{Bmatrix} + \\ \begin{pmatrix} k_3 & 0 & 0 \\ 0 & k_m(t) & k_{12} \\ 0 & k_{12} & k_2 + k_{12} \end{pmatrix} \begin{Bmatrix} x \\ u \\ v \end{Bmatrix} + \begin{pmatrix} 0 & 0 & 0 \\ 0 & s(t) & 0 \\ 0 & 0 & 0 \end{pmatrix} \begin{Bmatrix} x^3 \\ u^3 \\ v^3 \end{Bmatrix} = \begin{Bmatrix} 0 \\ f_1 \\ f_2 \end{Bmatrix}. \qquad (2)$$

Here, $k_m(t)$ comprises a linear stiffness $k_1 + k_{12}$ and a dual frequency harmonic term as follows:

$$k_m(t) = (k_1 + k_{12}) + k_1 \left( \alpha_a \cos(\omega_a t + \varphi_a) + \alpha_b \cos(\omega_b t + \varphi_b) \right). \qquad (3)$$

The parameters $\alpha_\bullet$ define the stiffness modulation magnitude of $k_1$ at the frequencies $\omega_a$ and $\omega_b$ with the appropriate phase shifts $\varphi_\bullet$.

## 3.2 Governing EOM in modal coordinates

The coupling of the EOM is reduced if the undamped linear system mass normalized modes are used to transform the coordinates. First, the mass normalized modes of the following system are computed:

$$\begin{pmatrix} \sum m_i & m_1 & -m_2 \\ m_1 & m_1 & 0 \\ -m_2 & 0 & m_2 \end{pmatrix} \ddot{\mathbf{q}} + \begin{pmatrix} k_3 & 0 & 0 \\ 0 & k_1 + k_{12} & k_{12} \\ 0 & k_{12} & k_2 \end{pmatrix} \mathbf{q} = \mathbf{0}, \qquad \mathbf{q} = \begin{Bmatrix} x \\ u \\ v \end{Bmatrix}. \qquad (4)$$

Using these normal modes, a modal matrix is defined as follows:

$$\boldsymbol{\Phi} = \begin{pmatrix} \boldsymbol{\phi}_1 & \boldsymbol{\phi}_2 & \boldsymbol{\phi}_3 \end{pmatrix}. \qquad (5)$$

Here, $\boldsymbol{\phi}_\bullet$ are the normal modes, and the coordinate transformation is defined as:

$$\mathbf{q} = \varepsilon^\alpha \boldsymbol{\Phi} \boldsymbol{\eta}, \quad \varepsilon \sim O(\hat{\zeta}), \qquad (6)$$

where $\varepsilon$ is a measure of the modal damping, and is assumed small:

$$0 \leq \varepsilon \ll 1. \qquad (7)$$

Plugging Eq.(6) into Eq.(2) and pre-multiplying it by $\boldsymbol{\Phi}^T$, leads to the following EOM in matrix notation:

$$\varepsilon^{\alpha}\mathbf{\Phi}^{T}\mathbf{M}\mathbf{\Phi}\ddot{\mathbf{\eta}}+\varepsilon^{\alpha}\mathbf{\Phi}^{T}\mathbf{C}\mathbf{\Phi}\dot{\mathbf{\eta}}+\varepsilon^{\alpha}\mathbf{\Phi}^{T}\mathbf{K}\mathbf{\Phi}\mathbf{\eta}+\varepsilon^{\alpha}\mathbf{\Phi}^{T}\mathbf{K}(t)\mathbf{\Phi}\mathbf{\eta}+\mathbf{\Phi}^{T}\mathbf{S}\left(\varepsilon^{\alpha}\mathbf{\Phi}\mathbf{\eta}\right)^{3}=\mathbf{\Phi}^{T}\mathbf{Q},$$

$$\mathbf{M}=\begin{pmatrix}\sum m_{i} & m_{1} & -m_{2}\\ m_{1} & m_{1} & 0\\ -m_{2} & 0 & m_{2}\end{pmatrix},\quad \mathbf{C}=\begin{pmatrix}c_{3} & 0 & 0\\ 0 & c_{1}+c_{12} & c_{12}\\ 0 & c_{12} & c_{2}+c_{12}\end{pmatrix},\quad \mathbf{S}=\begin{pmatrix}0 & 0 & 0\\ 0 & s(t) & 0\\ 0 & 0 & 0\end{pmatrix},\quad \mathbf{Q}=\begin{Bmatrix}0\\ f_{1}\\ f_{2}\end{Bmatrix}, \tag{8}$$

$$\mathbf{K}=\begin{pmatrix}k_{3} & 0 & 0\\ 0 & k_{1}+k_{12} & k_{12}\\ 0 & k_{12} & k_{2}\end{pmatrix},\quad \mathbf{K}(t)=\begin{pmatrix}0 & 0 & 0\\ 0 & k_{1}\begin{pmatrix}\alpha_{a}\cos(\omega_{a}t+\varphi_{a})+\\ +\alpha_{b}\cos(\omega_{b}t+\varphi_{b})\end{pmatrix} & 0\\ 0 & 0 & 0\end{pmatrix}.$$

Because the modal transformation was employed, the equations simplify to

$$\mathbf{I}\ddot{\mathbf{\eta}}+2\hat{\mathbf{\zeta}}\mathbf{\Omega}_{n}\dot{\mathbf{\eta}}+\mathbf{\Omega}_{n}^{2}\mathbf{\eta}+\mathbf{\Phi}^{T}\mathbf{K}(t)\mathbf{\Phi}\mathbf{\eta}+\varepsilon^{2\alpha}\mathbf{\Phi}^{T}\mathbf{S}(\mathbf{\Phi}\mathbf{\eta})^{3}=\varepsilon^{-\alpha}\mathbf{\Phi}^{T}\mathbf{Q},$$

$$\hat{\mathbf{\zeta}}=\begin{pmatrix}\hat{\zeta}_{1} & 0 & 0\\ 0 & \hat{\zeta}_{2} & 0\\ 0 & 0 & \hat{\zeta}_{3}\end{pmatrix},\quad \mathbf{\Omega}_{n}=\begin{pmatrix}\omega_{1} & 0 & 0\\ 0 & \omega_{2} & 0\\ 0 & 0 & \omega_{3}\end{pmatrix}. \tag{9}$$

Next, a dimensionless time is defined

$$\tau=\tilde{\omega}t, \tag{10}$$

where $\tilde{\omega}$ is the typical response frequency. Eliminating $\tau$ form Eq.(8) through Eq.(10) and dividing by $\tilde{\omega}^{2}$ leads to the scaled governing EOM.

$$\mathbf{I}\mathbf{\eta}''+\mathbf{\chi}^{2}\mathbf{\eta}=\mathbf{P}-\varepsilon\left(2\mathbf{\zeta}\mathbf{\chi}\mathbf{\eta}'+\mathbf{K}_{\gamma}(\tau)\mathbf{\eta}+\mathbf{\kappa}\tilde{\mathbf{\eta}}\right),$$

$$\mathbf{\chi}=\begin{pmatrix}\chi_{1} & 0 & 0\\ 0 & \chi_{2} & 0\\ 0 & 0 & \chi_{3}\end{pmatrix},\quad \chi_{\bullet}=\frac{\omega_{\bullet}}{\tilde{\omega}}, \tag{11}$$

Where $\bullet'$ stands for $\partial/\partial\tau$, and $\mathbf{\kappa}$ and $\tilde{\mathbf{\eta}}$ are given in Eq.(A.1).

Additionally, light damping, weak pumping and light nonlinearity are assumed:

$$\begin{aligned}\hat{\zeta}_{\bullet}&=\varepsilon\zeta_{\bullet} & \rightarrow & \quad\hat{\mathbf{\zeta}}=\varepsilon\mathbf{\zeta},\\ \alpha_{\bullet}&=\varepsilon\gamma_{\bullet} & \rightarrow & \quad\frac{1}{\tilde{\omega}^{2}}\mathbf{\Phi}^{T}\mathbf{K}(\tau)\mathbf{\Phi}=\varepsilon\mathbf{K}_{\gamma}(\tau),\\ s&=\varepsilon^{1-2\alpha}\kappa & \rightarrow & \quad\frac{1}{\tilde{\omega}^{2}}\varepsilon^{2\alpha}\mathbf{\Phi}^{T}\mathbf{S}(\mathbf{\Phi}\mathbf{\eta})^{3}=\varepsilon\mathbf{\kappa}\tilde{\mathbf{\eta}},\\ \mathbf{P}&=\frac{1}{\tilde{\omega}^{2}}\varepsilon^{-\alpha}\mathbf{\Phi}^{T}\mathbf{Q}.\end{aligned} \tag{12}$$

# 4 Analytical solutions using the method of multiple scales

In this section, the method of multiple scales [19] is used to derive the analytical solution of the governing scaled EOM for three cases. Therefore, the following solution is assumed:

$$\mathbf{\eta}(\varepsilon,\tau) = \mathbf{\eta}_0(\tau_0, \tau_1) + \varepsilon \mathbf{\eta}_1(\tau_0, \tau_1). \tag{13}$$

The solution comprises two spatial scales, $\mathbf{\eta}_0$ and $\mathbf{\eta}_1$, and two time scales, $\tau_i = \varepsilon^i \tau$, $i = 0, 1$. Plugging Eq.(13) into Eq.(11) and collecting terms of the same order of $\varepsilon$ lead to the following ODEs:

$\varepsilon^0$:
$$D_0^2 \mathbf{\eta}_0 + \chi^2 \mathbf{\eta}_0 = \mathbf{P}, \tag{14}$$

$\varepsilon^1$:
$$D_0^2 \mathbf{\eta}_1 + \chi^2 \mathbf{\eta}_1 = -\left(2 D_0 D_1 \mathbf{\eta}_0 + 2\zeta\chi D_0 \mathbf{\eta}_0 + \mathbf{K}_\gamma(\tau)\mathbf{\eta}_0 + \kappa \tilde{\mathbf{\eta}}_0\right). \tag{15}$$

Here, the notation $\partial/\partial\tau_\bullet \equiv D_\bullet$ was adopted.

First, the zeroth order equation is solved, where it is assumed that the external forces are harmonic with two distinct frequencies as follows:

$$f_1(\tau) = f_1 \cos(\chi_{r_1}\tau + \varphi_1), \quad f_2(\tau) = f_2 \cos(\chi_{r_2}\tau + \varphi_2). \tag{16}$$

Hence, the solution of Eq.(14) in complex form is:

$$\eta_{\bullet 0} = A_\bullet(\tau_1)e^{i\chi_\bullet \tau} + \Lambda_{\bullet 1}e^{i(\chi_{r_1}\tau + \varphi_1)} + \Lambda_{\bullet 2}e^{i(\chi_{r_2}\tau + \varphi_2)} + \text{CC}, \quad \bullet = 1, 2, 3$$

$$\Lambda_{\bullet 1} = \frac{\varepsilon^{-\alpha}}{2\tilde{\omega}^2} \frac{\Phi_{2\bullet} f_1}{\chi_\bullet^2 - \chi_{r_1}^2}, \quad \Lambda_{\bullet 2} = \frac{\varepsilon^{-\alpha}}{2\tilde{\omega}^2} \frac{\Phi_{3\bullet} f_2}{\chi_\bullet^2 - \chi_{r_2}^2}, \tag{17}$$

where, CC stands for complex conjugate of the preceding terms. The terms $A_\bullet(\tau_1)$ are computed from Eq.(15) once $\eta_{\bullet 0}$ have been substituted. The pumping frequencies has a major influence on the solution, particularly on the terms $A_\bullet(\tau_1)$, which are the amplitudes of the normal modes near the appropriate natural frequency.

From this point on it is assumed that the system is excited by external forces which are much slower than the first natural frequency $(\omega_{r_\bullet} \ll \omega_1 \Leftrightarrow \chi_{r_\bullet} \ll \chi_1)$, while being pumped by a dual frequency signal ($\omega_a$ and $\omega_b$). Additionally, it is assumed that the pumping frequencies can be tuned such that most of the response energy is directed to either one of the natural frequencies by tuning them per one of the three cases below:

$$\begin{array}{ll}
(1) & \chi_a \approx 2\chi_1, \quad \chi_b \approx \chi_1 - \chi_r \\
(2) & \chi_a \approx 2\chi_2, \quad \chi_b \approx \chi_2 - \chi_r \quad \text{while} \quad \chi_{r_1} = \chi_{r_2} \equiv \chi_r = \delta\chi_1, \quad 0 < \delta \ll 1. \\
(3) & \chi_a \approx 2\chi_3, \quad \chi_b \approx \chi_3 - \chi_r
\end{array} \tag{18}$$

## 4.1   Case 1 - First mode excitation

To transfer energy from one the of external forces to the first mode, hence amplify it; the pumping frequencies are tuned according to the first condition in Eq.(18):

$$\tilde{\omega} = \omega_1 \quad \Rightarrow \quad \chi_1 = 1, \quad \chi_r = \delta, \quad \chi_b \approx (1-\delta), \quad \chi_a \approx 2. \tag{19}$$

Two detuning parameters $\sigma_1$ and $\sigma_2$ are used to define the pumping frequencies:

$$\chi_b + \chi_r = \chi_1 + \varepsilon\sigma_1, \quad \chi_a = 2\chi_1 + \varepsilon\sigma_2. \tag{20}$$

Plugging Eq.(17) and Eq.(20) into Eq.(15), the following terms leading to secular terms arise and need to be nullified:

$\underline{\chi_1}:$
$$\mathrm{i}2\chi_1 A_1' + \mathrm{i}2\zeta_1\chi_1^2 A_1 + \frac{3\kappa\Phi_{21}^2 A_1}{\tilde{\omega}^2}\begin{pmatrix} \Phi_{21}^2 A_1\bar{A}_1 + \\ 2\begin{pmatrix} H_1^2 + H_2^2 + \\ \Phi_{22}^2 A_2\bar{A}_2 + \Phi_{23}^2 A_3\bar{A}_3 \end{pmatrix} \end{pmatrix} +$$
$$+ \frac{\Phi_{21}k_1}{2\tilde{\omega}^2}\left(\gamma_b\left(H_1\mathrm{e}^{\mathrm{i}\varphi_1} + H_2\mathrm{e}^{\mathrm{i}\varphi_2}\right)\mathrm{e}^{\mathrm{i}(\varphi_b+\sigma_1\tau_1)} + \gamma_a\Phi_{21}\bar{A}_1\mathrm{e}^{\mathrm{i}(\varphi_a+\sigma_2\tau_1)}\right) = 0 \tag{21}$$

$\underline{\chi_2}:$
$$\mathrm{i}2\chi_2 A_2' + \mathrm{i}2\zeta_2\chi_2^2 A_2 + \frac{3\kappa\Phi_{22}^2 A_2}{\tilde{\omega}^2}\begin{pmatrix} \Phi_{22}^2 A_2\bar{A}_2 + \\ 2\begin{pmatrix} H_1^2 + H_2^2 + \\ \Phi_{21}^2 A_1\bar{A}_1 + \Phi_{23}^2 A_3\bar{A}_3 \end{pmatrix} \end{pmatrix} = 0 \tag{22}$$

$\underline{\chi_3}:$
$$\mathrm{i}2\chi_3 A_3' + \mathrm{i}2\zeta_3\chi_3^2 A_3 + \frac{3\kappa\Phi_{23}^2 A_3}{\tilde{\omega}^2}\begin{pmatrix} \Phi_{23}^2 A_3\bar{A}_3 + \\ 2\begin{pmatrix} H_1^2 + H_2^2 + \\ \Phi_{21}^2 A_1\bar{A}_1 + \Phi_{22}^2 A_2\bar{A}_2 \end{pmatrix} \end{pmatrix} = 0 \tag{23}$$

Here,
$$H_1 = \Lambda_{11}\Phi_{21} + \Lambda_{21}\Phi_{22} + \Lambda_{31}\Phi_{23}, \quad H_2 = \Lambda_{12}\Phi_{21} + \Lambda_{22}\Phi_{22} + \Lambda_{32}\Phi_{23}. \tag{24}$$

Transforming $A_\bullet$ to polar form:

$$A_\bullet(\tau_1) = \frac{1}{2}a_\bullet(\tau_1)\mathrm{e}^{\mathrm{i}\phi_\bullet(\tau_1)}, \tag{25}$$

and separating Eq.(21) to Eq.(23) to real and imaginary parts leads to:

$$\underline{\chi_1-\text{Re}}: a_1\phi_1' = \frac{3\kappa\Phi_{21}^2 a_1}{8\tilde{\omega}^2\chi_1}\left(\Phi_{21}^2 a_1^2 + 2\left(4\left(H_1^2+H_2^2\right)+\Phi_{22}^2 a_2^2+\Phi_{23}^2 a_3^2\right)\right)+$$

$$+\frac{\Phi_{21}k_1}{4\tilde{\omega}^2\chi_1}\begin{pmatrix}2\gamma_b\begin{pmatrix}H_1\cos(\varphi_1+\varphi_b+\sigma_1\tau_1-\phi_1)+\\H_2\cos(\varphi_2+\varphi_b+\sigma_1\tau_1-\phi_1)\end{pmatrix}+\\\gamma_a\Phi_{21}a_1\cos(\varphi_a+\sigma_2\tau_1-2\phi_1)\end{pmatrix},$$

$$\underline{\chi_1-\text{Im}}: a_1' = -\zeta_1\chi_1 a_1 - \frac{\Phi_{21}k_1}{4\tilde{\omega}^2\chi_1}\begin{pmatrix}2\gamma_b\begin{pmatrix}H_1\sin(\varphi_1+\varphi_b+\sigma_1\tau_1-\phi_1)+\\H_2\sin(\varphi_2+\varphi_b+\sigma_1\tau_1-\phi_1)\end{pmatrix}+\\\gamma_a\Phi_{21}a_1\sin(\varphi_a+\sigma_2\tau_1-2\phi_1)\end{pmatrix},$$

$$\underline{\chi_2-\text{Re}}: a_2\phi_2' = \frac{3\kappa\Phi_{22}^2 a_2}{8\tilde{\omega}^2\chi_2}\left(\Phi_{22}^2 a_2^2 + 2\left(4\left(H_1^2+H_2^2\right)+\Phi_{21}^2 a_1^2+\Phi_{23}^2 a_3^2\right)\right), \quad (26)$$

$$\underline{\chi_2-\text{Im}}: a_2' = -\zeta_2\chi_2 a_2,$$

$$\underline{\chi_3-\text{Re}}: a_3\phi_3' = \frac{3\kappa\Phi_{23}^2 a_3}{8\tilde{\omega}^2\chi_3}\left(\Phi_{23}^2 a_3^2 + 2\left(4\left(H_1^2+H_2^2\right)+\Phi_{21}^2 a_1^2+\Phi_{22}^2 a_2^2\right)\right),$$

$$\underline{\chi_3-\text{Im}}: a_3' = -\zeta_3\chi_3 a_3.$$

To transform the ODE system to an autonomous form two new variables are introduced:

$$\psi_1 = \sigma_1\tau_1 - \phi_1, \quad \psi_2 = \sigma_2\tau_1 - 2\phi_1. \quad (27)$$

Differentiating $\psi_1$ and $\psi_2$ with respect to $\tau_1$ and equating them to zero, as one is interested in the steady-state solution, leads to $\sigma_1 = \sigma_2/2$, therefore, two variables are defined:

$$\sigma_1 \equiv \sigma_1 = \frac{1}{2}\sigma_2, \quad \psi_1 \equiv \psi_1 = \frac{1}{2}\psi_2. \quad (28)$$

These lead to an algebraic relation between the external excitation and pumping frequencies:

$$\chi_a = 2(\chi_r + \chi_b) \Leftrightarrow \omega_a = 2(\omega_r + \omega_b). \quad (29)$$

Eliminating $\psi_1$ and $\psi_2$ from Eq. (28) through Eq.(27) and setting the derivatives to zeroes, as one seeks the steady-state solution, leads to trivial solutions for the amplitudes of the second and third modes, while for the first mode the remaining equations are:

$$\underline{\chi_1 - \text{Re}:} \qquad a_{10}\sigma_1 = \frac{3\kappa\Phi_{21}^2 a_{10}}{8\tilde{\omega}^2 \chi_1}\left(\Phi_{21}^2 a_{10}^2 + 8\left(H_1^2 + H_2^2\right)\right) +$$

$$\frac{\Phi_{21}k_1}{4\tilde{\omega}^2 \chi_1}\left(2\gamma_b\begin{pmatrix}H_1 \cos(\varphi_1 + \varphi_b + \psi_{10}) + \\ H_2 \cos(\varphi_2 + \varphi_b + \psi_{10})\end{pmatrix} + \\ \gamma_a \Phi_{21} a_{10} \cos(\varphi_a + 2\psi_{10})\right), \qquad (30)$$

$$\underline{\chi_1 - \text{Im}:} \qquad a_{10} = -\frac{\Phi_{21}k_1}{4\tilde{\omega}^2 \zeta_1 \chi_1^2}\left(2\gamma_b\begin{pmatrix}H_1 \sin(\varphi_1 + \varphi_b + \psi_{10}) + \\ H_2 \sin(\varphi_2 + \varphi_b + \psi_{10})\end{pmatrix} + \\ \gamma_a \Phi_{21} a_{10} \sin(\varphi_a + 2\psi_{10})\right).$$

Here, the subscript $\bullet_0$ represents steady-state.

From the imaginary equation, the first mode's amplitude at steady-state is

$$a_{10} = -\frac{2\Phi_{21}k_1\gamma_b\left(H_1 \sin(\varphi_1 + \varphi_b + \psi_{10}) + H_2 \sin(\varphi_2 + \varphi_b + \psi_{10})\right)}{4\tilde{\omega}^2 \zeta_1 \chi_1^2 + \Phi_{21}^2 k_1 \gamma_a \sin(\varphi_a + 2\psi_{10})}. \qquad (31)$$

It can be seen from Eq.(31) that large amplitudes can be produced if the denominator approaches zero. For the denominator to approach zero, the following condition must be fulfilled:

$$\gamma_{\text{LTH},1} \triangleq \frac{4\omega_1^2 \zeta_1}{k_1 \Phi_{21}^2} \leq \gamma_a. \qquad (32)$$

The term on the left-hand side of Eq.(32) is termed the linear stability threshold. Additionally, it is noticeable that $a_{10}$ depends on the steady-state phase $\psi_{10}$ which is unknown. To compute $\psi_{10}$, $a_{10}$ is eliminated from the real equation of Eq.(30) through Eq.(31). This leads to a nonlinear transcendental equation from which the phase can be calculated as in Appendix A in [16]. Because the transcendental equation is nonlinear, it may have multiple solutions. The stability of each solution is evaluated in a standard procedure [11]. Once the phase is calculated it is substituted into Eq.(31) and the zero-order solution is given by:

$$\eta_1 \approx a_{10}\cos\left(\frac{\chi_a}{2}\tau - \psi_{10}\right) + 2\Lambda_{11}\cos(\chi_r\tau + \varphi_1) + 2\Lambda_{12}\cos(\chi_r\tau + \varphi_2),$$
$$\eta_2 \approx 2\Lambda_{21}\cos(\chi_r\tau + \varphi_1) + 2\Lambda_{22}\cos(\chi_r\tau + \varphi_2), \qquad (33)$$
$$\eta_3 \approx 2\Lambda_{31}\cos(\chi_r\tau + \varphi_1) + 2\Lambda_{32}\cos(\chi_r\tau + \varphi_2).$$

To better approximate the solution, terms form the first order are considered as well. The considered terms are at the first natural frequency $(\tilde{\omega} \approx \omega_a/2)$ and at the external force frequency. These next order terms are calculated by solving Eq.(A.2). Combining the zero-order solution with the partial first-order solution according to Eq.(13) leads to the following solution:

$$\eta_1 \approx a_{10} \cos\left(\frac{\chi_a}{2}\tau - \psi_{10}\right) + a_{11r}\cos(\chi_r\tau + \Psi_{11r}),$$

$$\eta_2 \approx a_{21} \cos\left(\frac{\chi_a}{2}\tau + \Psi_{21}\right) + a_{21r}\cos(\chi_r\tau + \Psi_{21r}), \quad (34)$$

$$\eta_3 \approx a_{31} \cos\left(\frac{\chi_a}{2}\tau + \Psi_{31}\right) + a_{31r}\cos(\chi_r\tau + \Psi_{31r}),$$

where the terms $a_{\bullet 1}$, $a_{\bullet 1r}$, $\Psi_{\bullet 1}$ and $\Psi_{\bullet 1r}$ are omitted for brevity.

## 4.2 Second and third mode excitation

In a similar manner, as in the previous section, energy can be transferred from the external forces to the second or third mode. Therefore, the pumping frequencies are tuned according to the second or third condition in Eq.(18):

$$\begin{aligned}\tilde{\omega} = \omega_2 &\Rightarrow \chi_2 = 1, \quad \chi_r = \delta\chi_1, \quad \chi_b \approx (1-\delta\chi_1), \quad \chi_a \approx 2, \\ \tilde{\omega} = \omega_3 &\Rightarrow \chi_3 = 1, \quad \chi_r = \delta\chi_1, \quad \chi_b \approx (1-\delta\chi_1), \quad \chi_a \approx 2.\end{aligned} \quad (35)$$

Following the same procedure for these parameter as in the previous section, leads to the same algebraic relation between the pumping and external force frequencies, Eq.(29). However, the linear threshold stabilities differ:

$$\gamma_{\text{LTH},2} \triangleq \frac{4\omega_2^2 \zeta_2}{k_1 \Phi_{22}^2}, \quad \gamma_{\text{LTH},3} \triangleq \frac{4\omega_3^2 \zeta_3}{k_1 \Phi_{23}^2}. \quad (36)$$

Nevertheless, the zero-order solution and the partial first-order solution can be derived and combined to produce a better approximation.

For the second natural frequency:

$$\eta_1 \approx a_{11} \cos\left(\frac{\chi_a}{2}\tau + \Psi_{11}\right) + a_{11r}\cos(\chi_r\tau + \Psi_{11r}),$$

$$\eta_2 \approx a_{20} \cos\left(\frac{\chi_a}{2}\tau - \psi_{20}\right) + a_{21r}\cos(\chi_r\tau + \Psi_{21r}), \quad (37)$$

$$\eta_3 \approx a_{31} \cos\left(\frac{\chi_a}{2}\tau + \Psi_{31}\right) + a_{31r}\cos(\chi_r\tau + \Psi_{31r}).$$

And for the third natural frequency:

$$\eta_1 \approx a_{11} \cos\left(\frac{\chi_a}{2}\tau + \Psi_{11}\right) + a_{11r}\cos(\chi_r\tau + \Psi_{11r}),$$

$$\eta_2 \approx a_{21} \cos\left(\frac{\chi_a}{2}\tau + \Psi_{21}\right) + a_{21r}\cos(\chi_r\tau + \Psi_{21r}), \quad (38)$$

$$\eta_3 \approx a_{30} \cos\left(\frac{\chi_a}{2}\tau - \psi_{30}\right) + a_{31r}\cos(\chi_r\tau + \Psi_{31r}).$$

where the terms $a_{\bullet 1}$, $a_{\bullet 1r}$, $\Psi_{\bullet 1}$ and $\Psi_{\bullet 1r}$ differ from Eq.(34), Eq.(37) and Eq.(38), and are omitted for brevity.

## 5 Intermediate summary and conclusions

In the previous section, the responses at steady-state were analytically approximated using the method of multiple scales. Additionally, the different solution stability was evaluated. It was found that to parametrically excite a certain mode, different magnitudes of pumping are needed to overcome the linear stability threshold.

$$\gamma_{\text{LTH},1} \triangleq \frac{4\omega_1^2 \zeta_1}{k_1 \Phi_{21}^2}, \quad \gamma_{\text{LTH},2} \triangleq \frac{4\omega_2^2 \zeta_2}{k_1 \Phi_{22}^2}, \quad \gamma_{\text{LTH},3} \triangleq \frac{4\omega_3^2 \zeta_3}{k_1 \Phi_{23}^2}. \tag{39}$$

From Eq.(39), the parameters governing the magnitude of the LTH depend on the topology, pumping position and modal damping. The topology determines the natural frequencies $\omega_\bullet$ and the mode shapes $\mathbf{\Phi}$. In this system, the pumping is done by modulating the linear stiffness $k_1$, which connects masses $m_1$ and $m_3$. The mases' relative position is described by the coordinate $u(t)$, which suits the terms $\Phi_{2\bullet}$.

To excite the second and third modes, the system parameters were designed such that the appropriate LTH obeys $\gamma_{\text{LTH},\bullet} < 0.3/\varepsilon$. This way it is guaranteed that the stiffness is always positive. Additional constraints were posed on the topology to avoid internal resonances which arise due to the nonlinearity.

It can be deduced, that the LTHs are crucial design parameters one must consider prior to designing a MDOF parametric amplifier. Therefore, to design the following experimental system, a nonlinear optimization scheme was used to set the different system parameters: springs and masses, while the modal damping was assumed to be 3%.

## 6 Numerical and experimental verification

The experimental system was built according to the design parameters which were calculated using the nonlinear optimization scheme. In order that the analytical and numerical models fit the experimental system, its parameters were estimated experimentally. Once the models were in accordance, numerical verification and experimental validation were carried out.

Several situations were studied with two different topologies and are listed in Table 1. For all cases, the ability to amplify the input force and sense it (i.e., amplitude and phase) were investigated.

| Case study | Topology | External force position | Excited normal mode |
|---|---|---|---|
| 1 | without added link $k_{12} = 0,\ c_{12} = 0$ | $u \to f_1$ | Mode 2: $\omega_2,\ \eta_2$ |
| 2 |  | $v \to f_2$ |  |
| 3 | with added link $k_{12} \neq 0,\ c_{12} \neq 0$ | $u \to f_1$ | Mode 3: $\omega_3,\ \eta_3$ |
| 4 |  | $v \to f_2$ |  |

Table 1. The different case studies.

## 6.1 Experimental system identification

To identify the experimental system parameters: masses, linear stiffness, and damping several methods were used. First, the sensors were calibrated and the mild nonlinearity was compensated by model inversion, assumed to be a third order polynomial.

Afterwards, the modal parameters (i.e., natural frequencies, normal modes and modal damping) of the linear system were identified using the Structural Dynamics Toolbox (SDT) [20] by a nonlinear iterative procedure. To use the SDT, six frequency responses composing the transfer function matrix were computed by applying two distinct harmonic force vectors with increasing frequencies aka stepped sine. By doing so, the different elements of the transfer function matrix were computed at each frequency as follows:

$$\mathbf{H} = \begin{bmatrix} \mathbf{S}_1 & \mathbf{S}_2 \end{bmatrix} \begin{bmatrix} \mathbf{F}_1 & \mathbf{F}_2 \end{bmatrix}^{-1}. \tag{40}$$

Here $\mathbf{H} \in \mathbb{C}^{3\times 2}$ is the transfer function matrix, $\mathbf{S}_i \in \mathbb{C}^{3\times 1}$ is the measured response vector due to the i$^{\text{th}}$ force vector $\mathbf{F}_i \in \mathbb{C}^{2\times 1}$. From the SDT, the natural frequencies, mode shapes and damping ratios were extract, by which the mass, stiffness and damping matrix were reconstructed according to Eq.(8). Following the aforementioned procedure, the resulting matrices were full in contrast to the model's matrices, Eq.(2). To force the reconstructed matrices topology according to the model, a nonlinear optimization problem was solved as described in Appendix B, and the estimated parameters for the first topology are:

$$\begin{aligned} & m_1 \approx 351.4 \text{ g}, \quad m_2 \approx 368.9 \text{ g}, \quad m_3 \approx 5.6473 \text{ kg}, \\ & k_1 \approx 1.06 \text{ kN m-1}, \quad k_2 \approx 1.05 \text{ kN m-1}, \quad k_3 \approx 10.04 \text{ kN m-1}. \end{aligned} \tag{41}$$

For the second topology, with the virtually added spring and dashpot, the identified parameters are:

$$\begin{aligned} & m_1 \approx 305.9 \text{ g}, \quad m_2 \approx 360.9 \text{ g}, \quad m_3 \approx 5.6575 \text{ g}, \\ & k_1 \approx 1.06 \text{ kN m-1}, \quad k_2 \approx 1.04 \text{ kN m-1}, \quad k_3 \approx 10.08 \text{ kN m-1}, \\ & k_{12} \approx 0.52 \text{ kN m-1}. \end{aligned} \tag{42}$$

The various mode shapes and appropriate natural frequencies of the two different topologies are shown in Figure 5 and Figure 6.

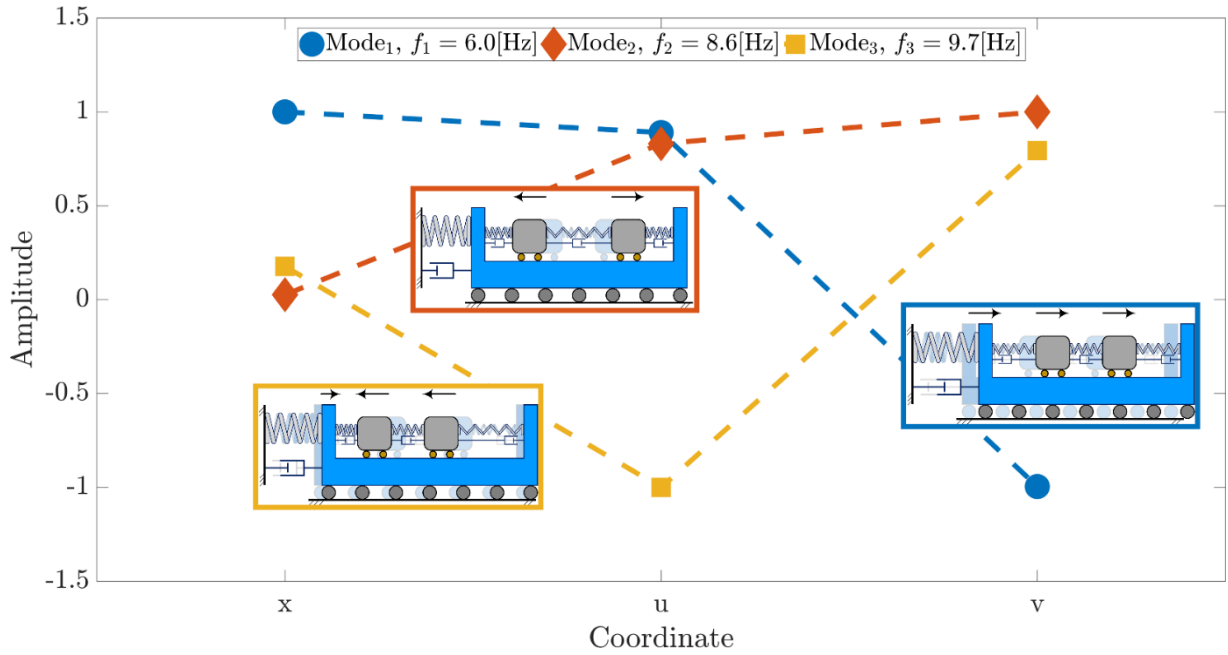

Figure 5. First topology estimated mode shapes and natural frequencies.

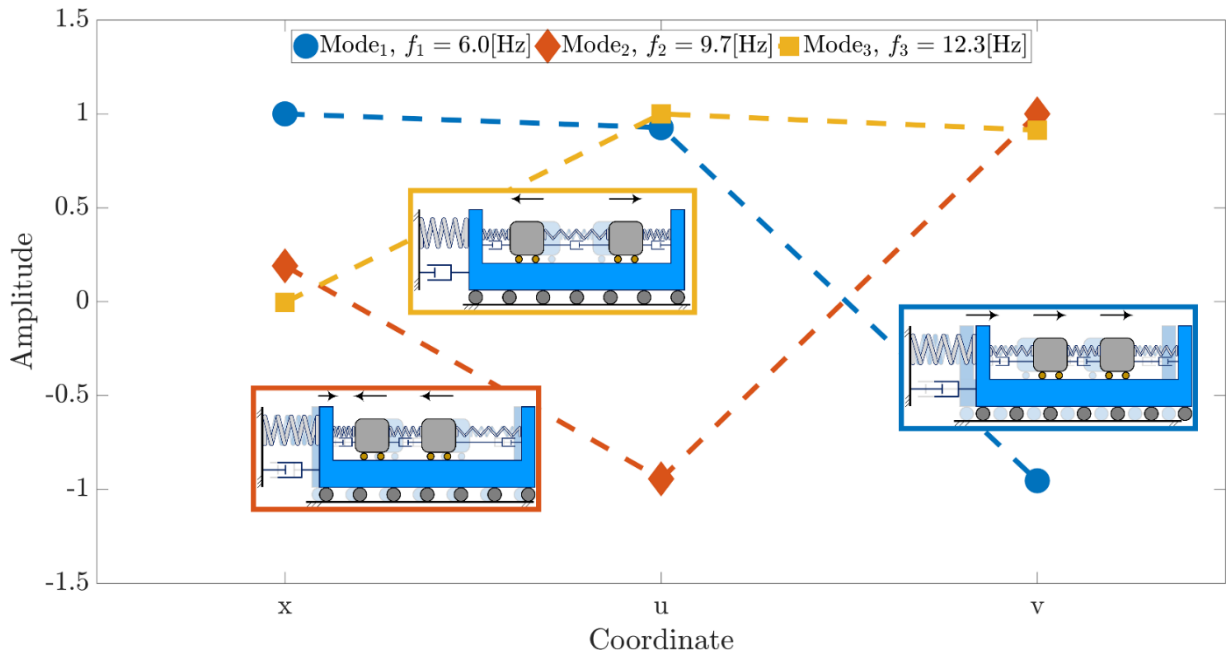

Figure 6. Second topology estimated mode shapes and natural frequencies.

## 6.2 Experiments, numerical simulations and analytical solutions

The ability to sense and amplify a single input was studied analytically, verified numerically, and validated experimentally. Moreover, the amplifier sensitivity to the input amplitude and phase were addressed. In the following subsections, first the ability to amplify the inputs using the DFPA scheme is studied by performing a frequency sweep, during which the input signal amplitude was 0.2 N and the frequency was 0.84 Hz. Then, the sensitivities to the input amplitude and phase were evaluated qualitatively.

### 6.2.1 Case study 1

In this case, the system was in its original configuration (i.e., first topology) and the force was applied between $m_1$ and $m_3$, parallel to the linear spring $k_1$ whose stiffness was modulated. To excite the second mode, the pumping frequencies were chosen according to the second condition in Eq.(18). The pumping magnitude $\gamma_a$ was set according to Eq.(39) as $\gamma_a = 1.02\gamma_{\text{LTH},2}$ to produce large amplitudes, and $\gamma_b$ as $1.05\gamma_{\text{LTH},2}$ to produce coupling between the input and output.

In figures 7 through 16, the amplitudes of the three modes close to the natural frequencies are depicted, according to Eq.(37) and Eq.(38). The amplitudes depicted in the figures were computed analytically, numerically and experimentally.

Figure 7 depicts a frequency scan, for which the input signal is $f_1$. In this case the input signal was amplified by producing large amplitudes of the second mode, $a_{20}$. The analytical, simulated and experimental results agree well up to ~8.7 Hz. Above this frequency, the amplitudes become relatively large, and the asymptotic model is no longer valid.

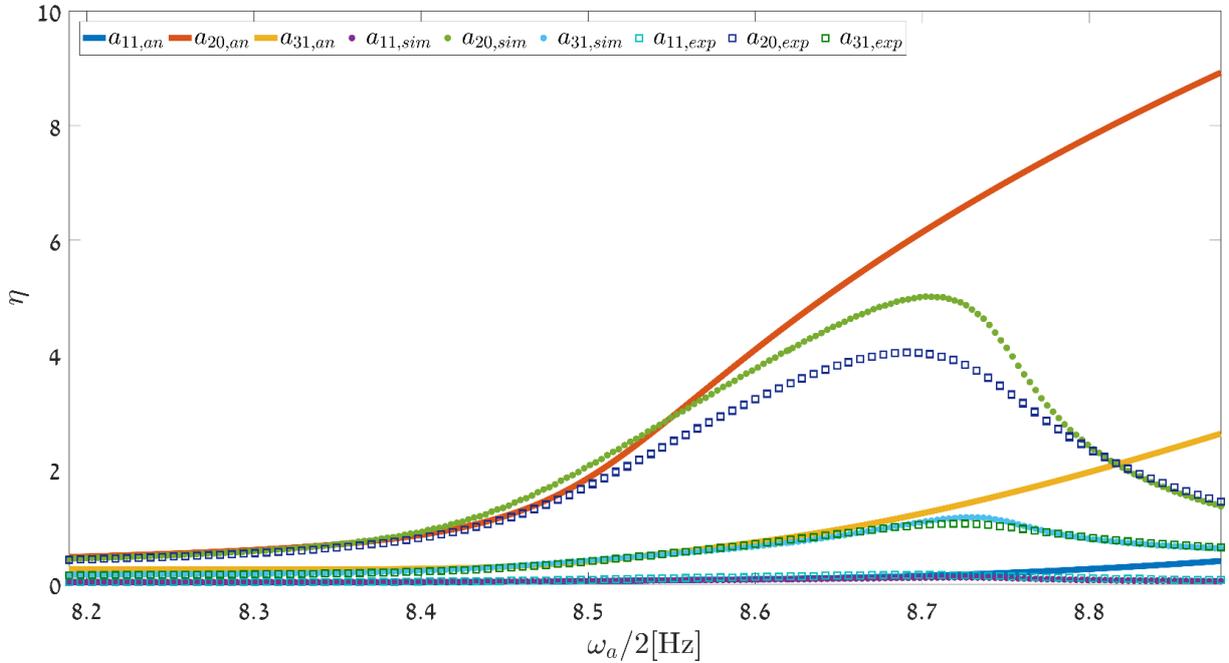

Figure 7. Topology 1 – frequency scan. Normal mode amplitudes vs. the pumping frequency while $f_1$=0.2 N and $f_2$=0. Continuous lines / circular markers / hollow square markers show analytical / simulated / experimental results.

Once it was established that the results agree, and the DFPA scheme works, hence the input is amplified, the response sensitivity to the input amplitude and phase were estimated qualitatively. In Figure 8, the various amplitudes vs. the input amplitude are shown, when the detuning parameter $\sigma$ was set as -1.39. One can witness that the trend of the experimental results resembles the analytical and simulated results, and even surpass them in terms of sensitivity, which is the slope of the curves. It can be deduced that as the input is weaker the sensitivity is higher, and that beyond a certain value ($f_1 \approx 0.15$ N for the experimental curve) the sensitivity approaches zero and even changes sign.

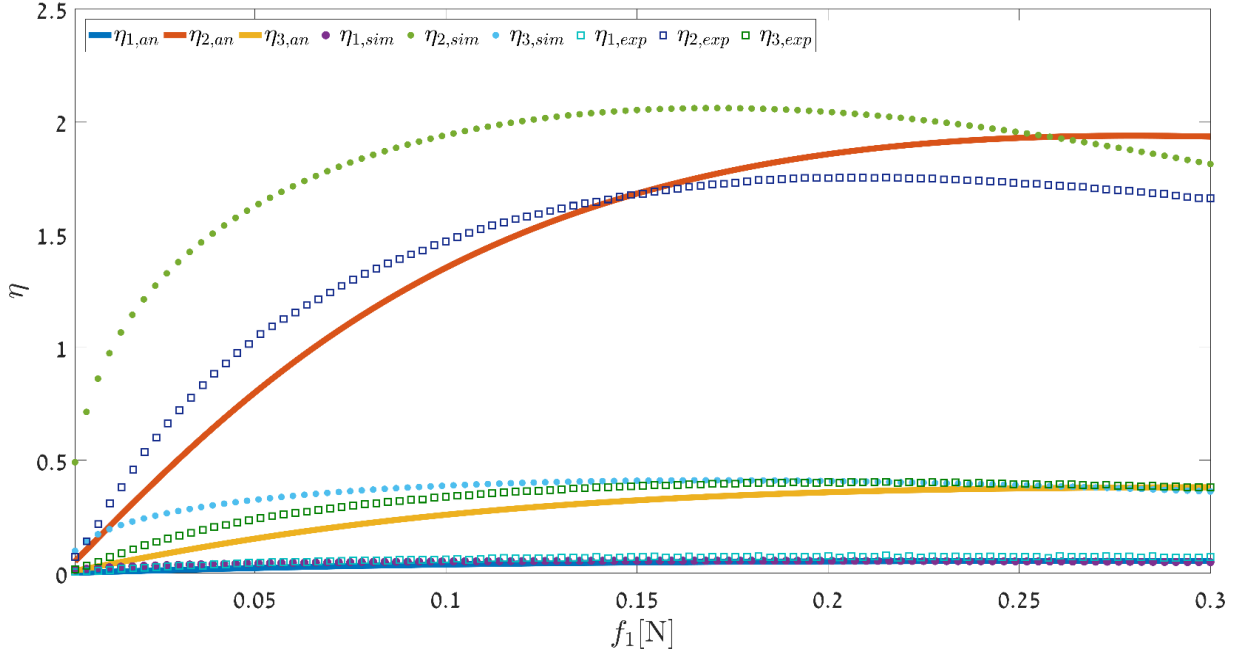

Figure 8. Topology 1 – normal mode amplitudes vs. $f_1$ amplitude at $\sigma$=-1.39, while $f_2$=0. Continuous lines / circular markers / hollow square markers show analytical / simulated / experimental results.

Next, the sensitivity to the input phase was evaluated by measuring the response vs. $\varphi_1$. The results are shown in Figure 9, and again the measurements and simulated results agree with the analytical ones. In contrast to linear systems where the input phase does not affect the response amplitude, in this case it does with a $\pi$ period.

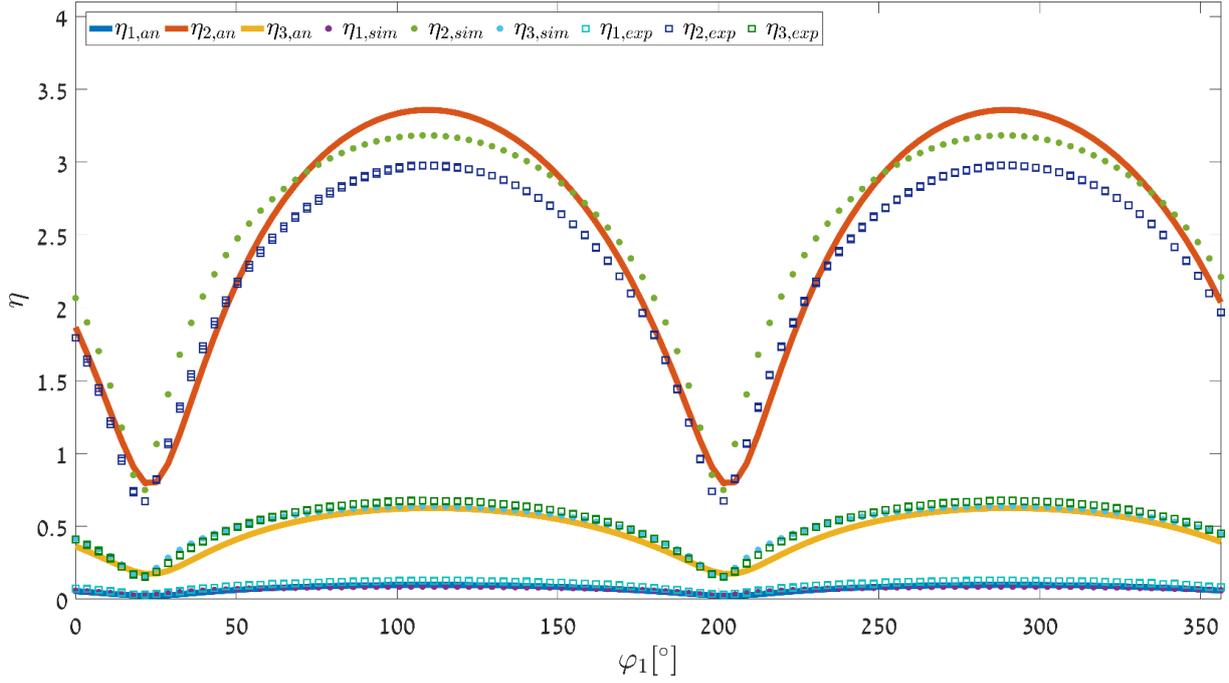

Figure 9. Topology 1 – normal mode amplitudes vs. $f_1$ phase, $\varphi_1$ at $\sigma=-1.39$, while $f_1=0.2$ N and $f_2=0$. Continuous lines / circular markers / hollow square markers show analytical / simulated / experimental results.

### 6.2.2 Case study 2

The topology and pumping parameters were tuned as in the first case study, however in this case the force is applied between $m_2$ and $m_3$ (i.e., $f_2$). The results are depicted in Figure 10, and in contrast to the first case the results do not agree with the model. During the experiment, some dynamics was observed, but it was rather random and incoherent. Different dynamics was observed in the simulated results indicating that some amplification had to take place, however according to the analytical model no amplification was supposed to be produced. The weak analytical solution, not visible in Figure 10, implies that it is impossible to amplify the input signal using the DFPA scheme. The dynamics observed in the simulated and measured results are probably due to principal parametric resonance, which alone cannot be used to amplify the input, although it produces large amplitudes. It may have occurred due to measurement noise during the experiment, and numerical errors in the simulation. Nevertheless, the produced amplitudes are smaller than the ones observed in Figure 7. Therefore, it can be argued that using this configuration the input signal $f_2$ cannot be amplified. Further explanation regarding the inability to amplify $f_2$ using this configuration is provided in the last section.

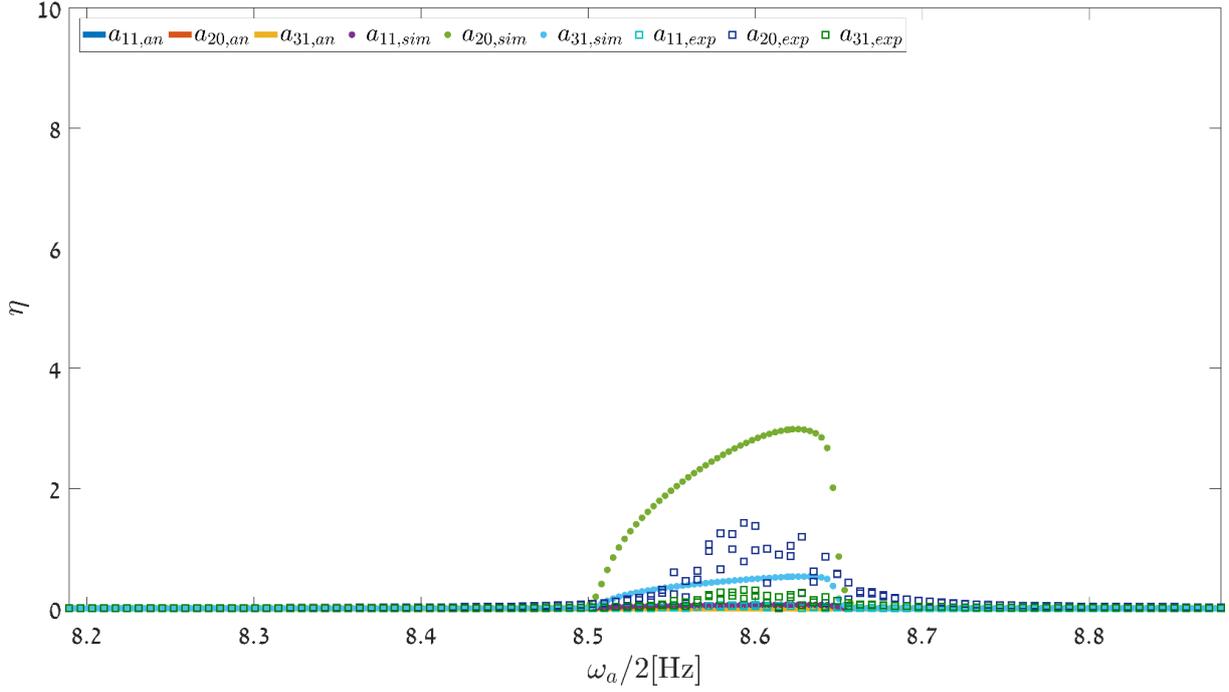

Figure 10. Topology 1 – frequency scan. Normal mode amplitudes vs. the pumping frequency while $f_1$=0 and $f_2$=0.2 N. Continuous lines / circular markers / hollow square markers show analytical / simulated / experimental results.

### 6.2.3 Case study 3

To overcome the inability to amplify the input signal $f_2$, it was suggested to modify the system's topology. Therefore, a coupling link between $m_1$ and $m_2$ comprising a linear spring $k_{12}$ and a linear viscous dashpot $c_{12}$ was added digitally by applying position and velocity related forces.

Prior to evaluating the ability to overcome the problem of amplifying $f_2$, the ability to amplify the signal $f_1$ with this new topology was verified. With this new topology and input signal $f_1$ applied between $m_1$ and $m_3$, the pumping frequencies were chosen according to the third condition in Eq.(18) to excite the third mode. This mode was chosen because it spatially matches the second mode of the first topology (see Figure 5 and Figure 6). The pumping magnitude $\gamma_a$ was set according to Eq.(39) as $\gamma_a = 1.02\gamma_{LTH,3}$ to produce large amplitudes, and $\gamma_b$ as $1.05\gamma_{LTH,3}$ to produce coupling between the input and output.

A frequency scan, for which the input signal is $f_1$ is depicted in Figure 11. In this case the input signal was amplified by producing large amplitudes of the third mode, $a_{30}$. The experimental results are slightly shifted (relative error $\approx 2.28\%$) to the left with respect to the analytical and simulated ones. This slight shift indicates minor errors in the parameters estimation, nevertheless the error is sufficiently small for the results to agree. This case emphasizes the importance of accurate system identification procedure, especially because the response to parametric excitation is very narrow banded.

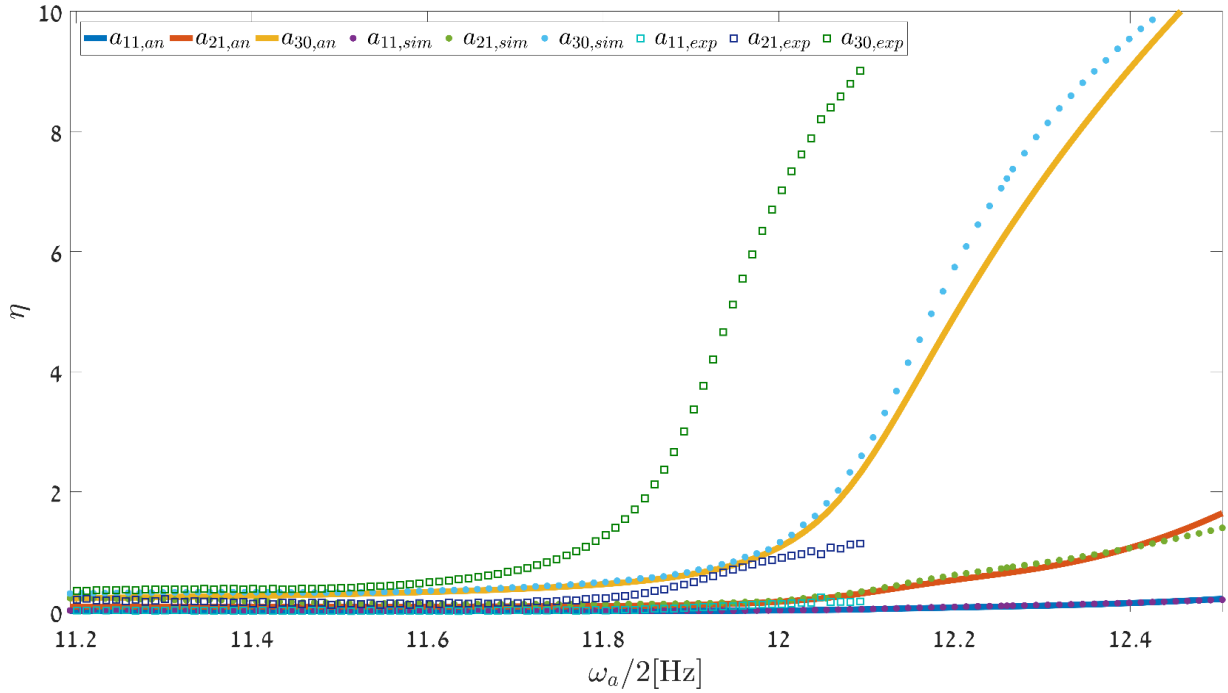

Figure 11. Topology 2 – frequency scan. Normal mode amplitudes vs. the pumping frequency while $f_1$=0.2 N and $f_2$=0. Continuous lines / circular markers / hollow square markers show analytical / simulated / experimental results.

As in the first case, the input was amplified and the results agree, hence the sensitivities were studied via two experiments. First the normal mode amplitudes vs. the input amplitude were computed and are shown in Figure 12, where the experiment was conducted with $\sigma_{\exp} = -3.27$, and the analytical and simulated results were computed for $\sigma_{an} = -1.43$. The detuning parameters differ to compensate for the slight shift observed in Figure 11. Then, the normal mode amplitudes vs. the input phase were computed and are shown in Figure 12, where the detuning parameters remained as in the latter experiment. In both cases the results agree, and it can be seen that the response is indeed sensitive with respect to the input amplitude and phase.

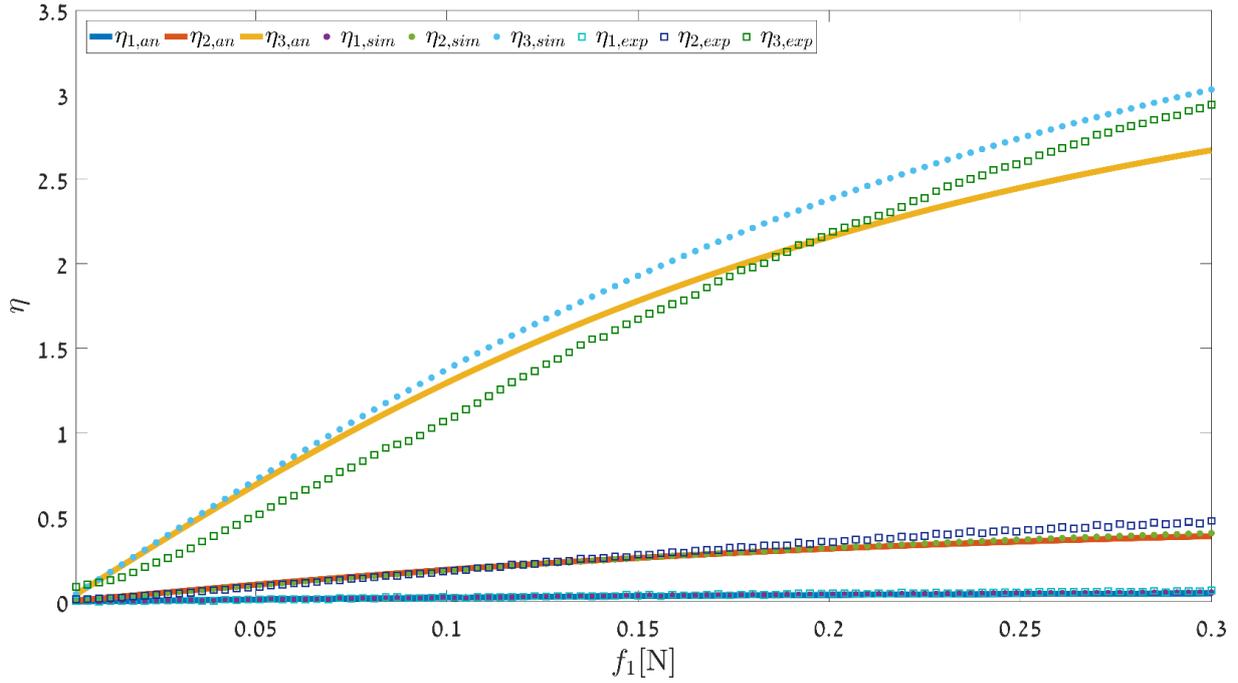

Figure 12. Topology 2 – normal mode amplitudes vs. $f_1$ amplitude at $\sigma_{exp}$=-3.27, $\sigma_{an}$=-1.43, while $f_2$=0. Continuous lines / circular markers / hollow square markers show analytical / simulated / experimental results.

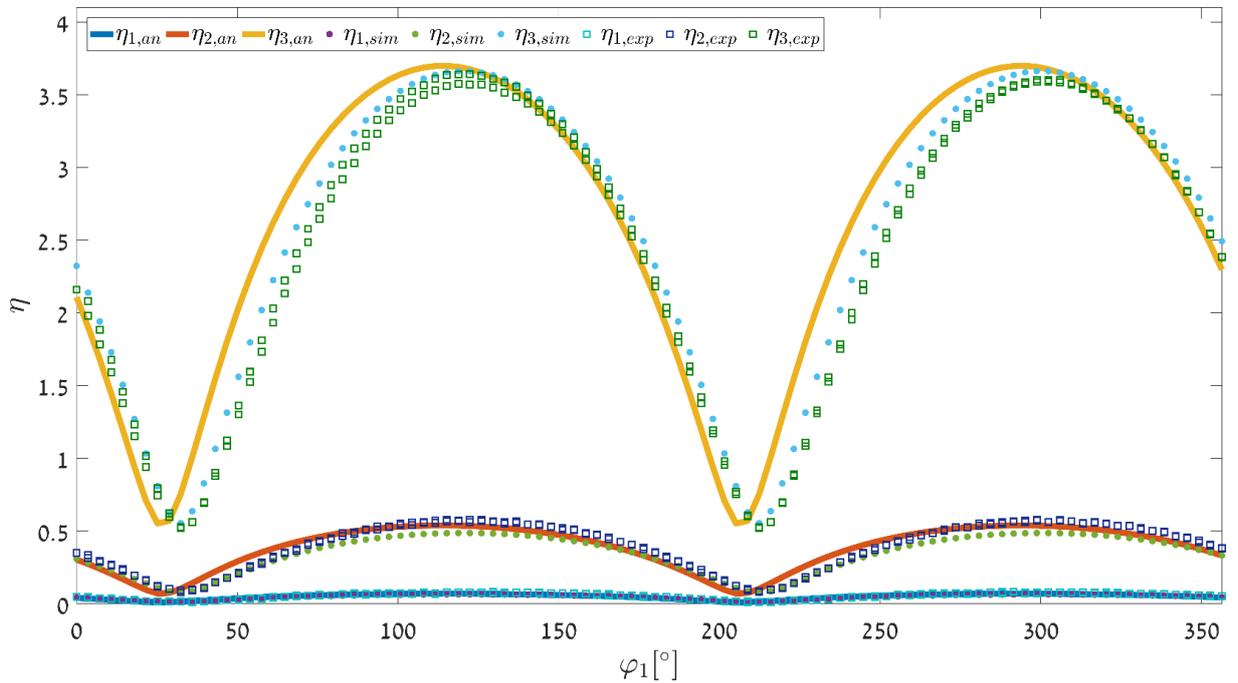

Figure 13. Topology 2 – normal mode amplitudes vs. $f_1$ phase, $\varphi_1$ at $\sigma_{exp}$=-3.27, $\sigma_{an}$=-1.43, while $f_1$=0.2 N and $f_2$=0. Continuous lines / circular markers / hollow square markers show analytical / simulated / experimental results.

### 6.2.4 Case study 4

With the second topology, it was shown in the Section 6.2.3 that the input $f_1$ can be amplified, therefore now the ability to amplify $f_2$ is evaluated by setting the pumping frequencies and magnitudes as in the previous case. However, in the following experiments the input signal was $f_2$, and the frequency scan is shown in Figure 14. As in the previous case, the input signal was amplified by producing large amplitudes of the third mode, $a_{30}$. In a similar manner to the results shown in Figure 11, the experimental result are slightly shifted (relative error $\approx 2.28\%$) to the left due to minor errors discussed previously. Yet, the error is sufficiently small for the results to agree.

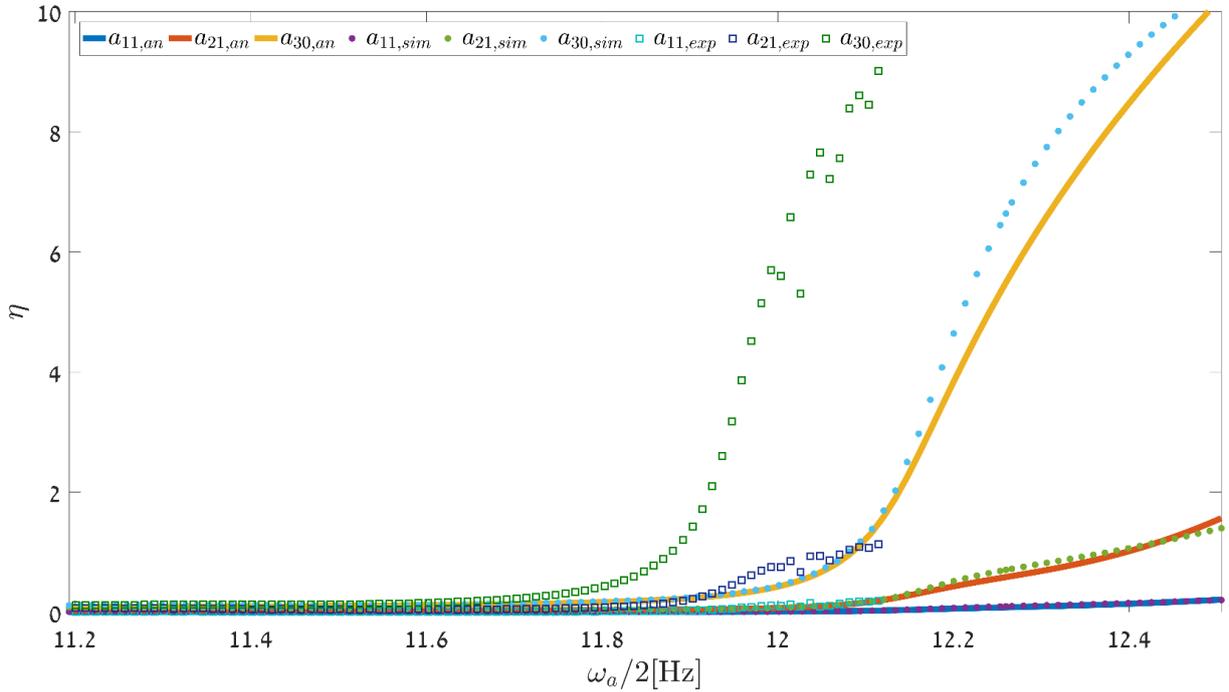

Figure 14. Topology 2 – frequency scan. Normal mode amplitudes vs. the pumping frequency while $f_1$=0 and $f_2$=0.2 N. Continuous lines / circular markers / hollow square markers show analytical / simulated / experimental results.

In this case, the input was amplified and the results agree, hence the sensitivities were studied via two experiments as before. The normal mode amplitudes vs. the input amplitude were computed and are shown in Figure 15, where the experiment was conducted with $\sigma_{exp} = -2.72$, and the analytical and simulated results were computed for $\sigma_{an} = -0.96$. As in Section 6.2.3, the detuning parameters differ to compensate for the slight shift observed in Figure 14. Then, the normal mode amplitudes vs. the input phase were computed and are shown in Figure 16, where the detuning parameters remained as in the latter experiment. In both cases, the results agree, and it can be seen that the response is indeed sensitive with respect to the input amplitude and phase.

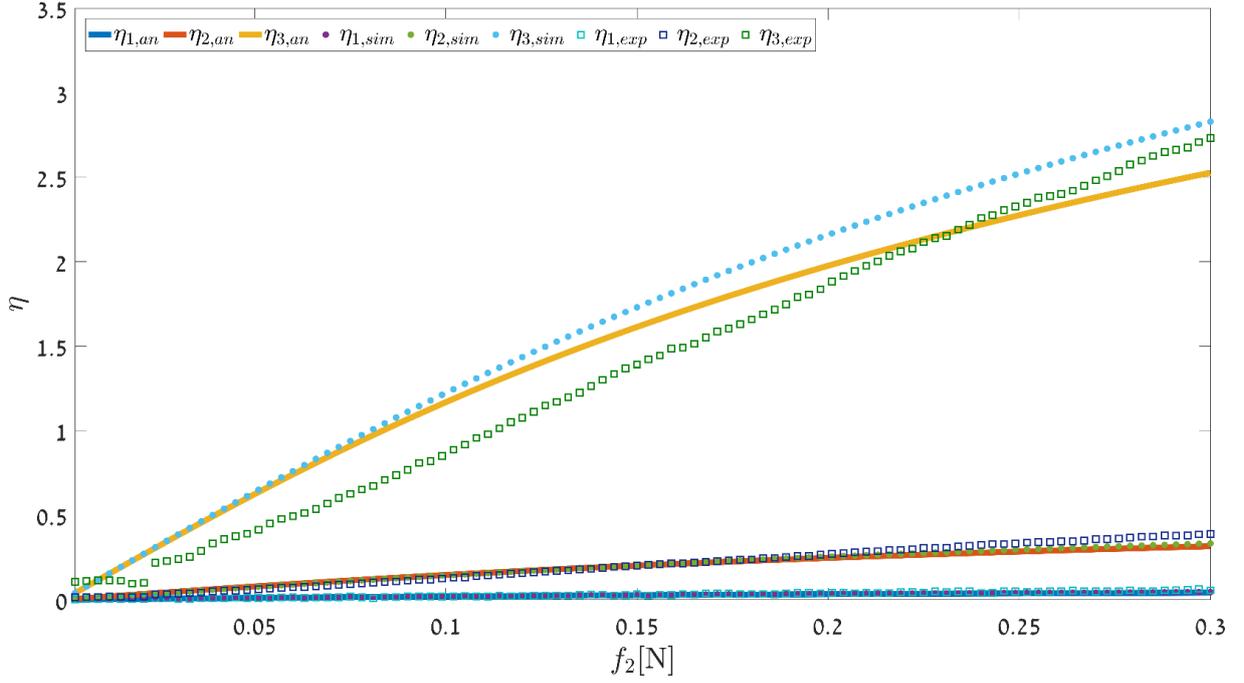

Figure 15. Topology 2 – normal mode amplitudes vs. $f_2$ amplitude at $\sigma_{exp}$=-2.72, $\sigma_{an}$=-0.96, while $f_1$=0 and $f_2$=0.2 N. Continuous lines / circular markers / hollow square markers show analytical / simulated / experimental results.

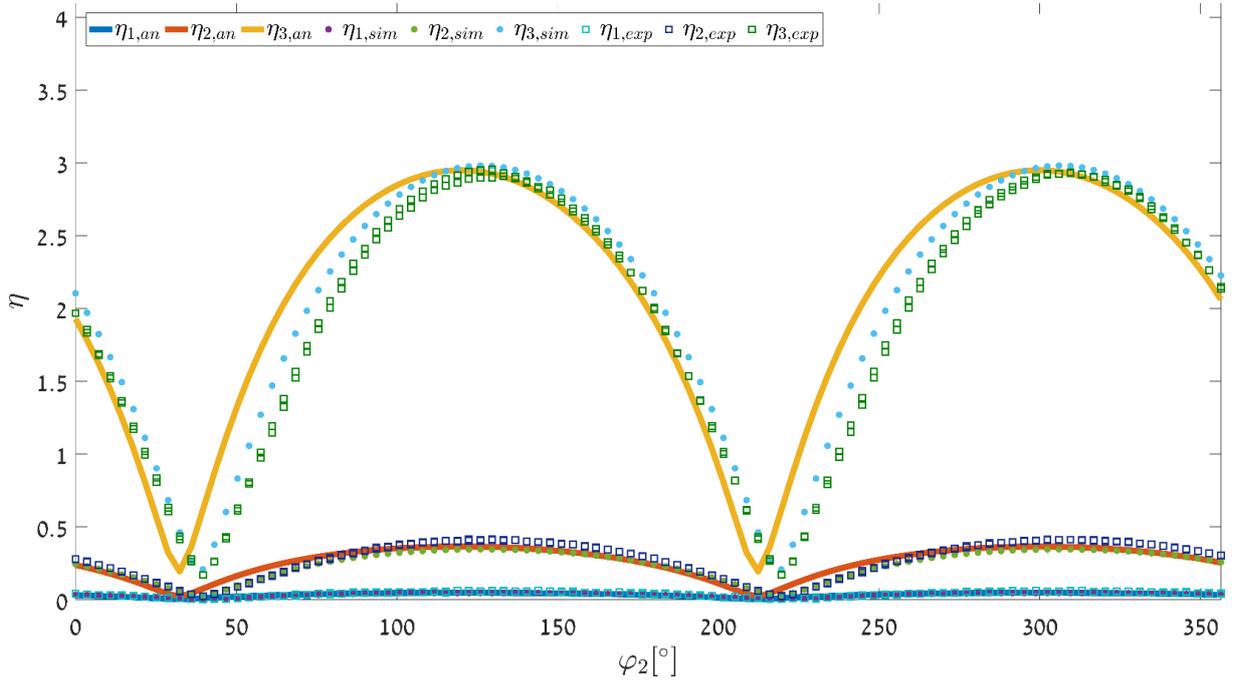

Figure 16. Topology 2 – normal mode amplitudes vs. $f_2$ phase, $\varphi_2$ at $\sigma_{exp}$=-2.72, $\sigma_{an}$=-096, while $f_1$=0 and $f_2$=0.2 N. Continuous lines / circular markers / hollow square markers show analytical / simulated / experimental results.

## 7. Conclusions

In this work the performance of a nonlinear, three degree of freedom DFPA with digitally tunable topology was studied. The presented experimental rig was carefully designed using a nonlinear optimization procedure to avoid unwanted dynamics, and to allow parametric excitation with relatively low pumping magnitudes. The criteria for the optimization are based on the asymptotic analytical solution provided in Section 4, among which are the linear stability thresholds. It was found that the latter depend on the amplifier topology and parametric excitation position.

Prior to conducting the experiments, a thorough system identification procedure was done to estimate the different system parameters. This procedure incorporated several stages: actuators calibration, nonlinear modal parameters estimation and a linear model update followed by a nonlinear model update. This stage proved to be of high importance as the experimental results agreed with the simulated and analytical results, especially because the response to parametrically excitation is very narrow banded. In the various cases presented, the response bandwidth was less than 1 Hz, which is about 1% of the appropriate natural frequency. Although the results agreed, evidence to the slight identification errors was observed when the second topology was used, in Figure 11 and Figure 14.

Once the model parameters were identified, the ability to sense and amplify two different inputs by projecting their influence on the amplifier normal modes was studied. Furthermore, once the input signal was indeed amplified, the sensitivity to its amplitude and phase were qualitatively studied. It was shown that when the first topology (i.e., without added stiffness link) was used, only the input $f_1$ could be amplified. To overcome the inability to amplify both inputs, the amplifier's topology was digitally modified by addition of a stiffness link connecting $m_1$ and $m_3$. Using the modified topology, it was shown that both inputs could be amplified, and the response was influenced by change in the inputs amplitude and phase. It is worth noting that the digital topology modification is not limited to proximate points of the structure and can be used to couple originally uncoupled structures.

The inability to amplify $f_2$ using the first topology, and the ability to amplify it using the second topology can be explained by examination of the analytic linear frequency responses of the system. The frequency response output is the $u$ coordinate, where the parametric excitation is done, and the input is either $f_1$ or $f_2$ as shown in Figure 17. The vertical black line depicts the frequency of the input signals, $\omega_r$. The frequency response of the first topology is shown with continues lines, and it is noticeable that the response due to $f_1$ is about four magnitudes larger than the response due to $f_2$. This means that more energy from the first input reaches the pumped spring than from the second input, which explains why only the first input was amplified. On the other hand, when the second topology was employed, shown by dotted lines in Figure 17, the response due to both inputs is of the same order of magnitude, hence both were amplified.

It can be deduced that if the system is pumped at twice the natural frequency with magnitude larger than the linear stability threshold principal parametric resonance occurs. However, to amplify the input signal enough energy from it should reach the pumped spring. Once these two conditions are met, the signal can be amplified, and its amplitude and phase affect the response amplitude.

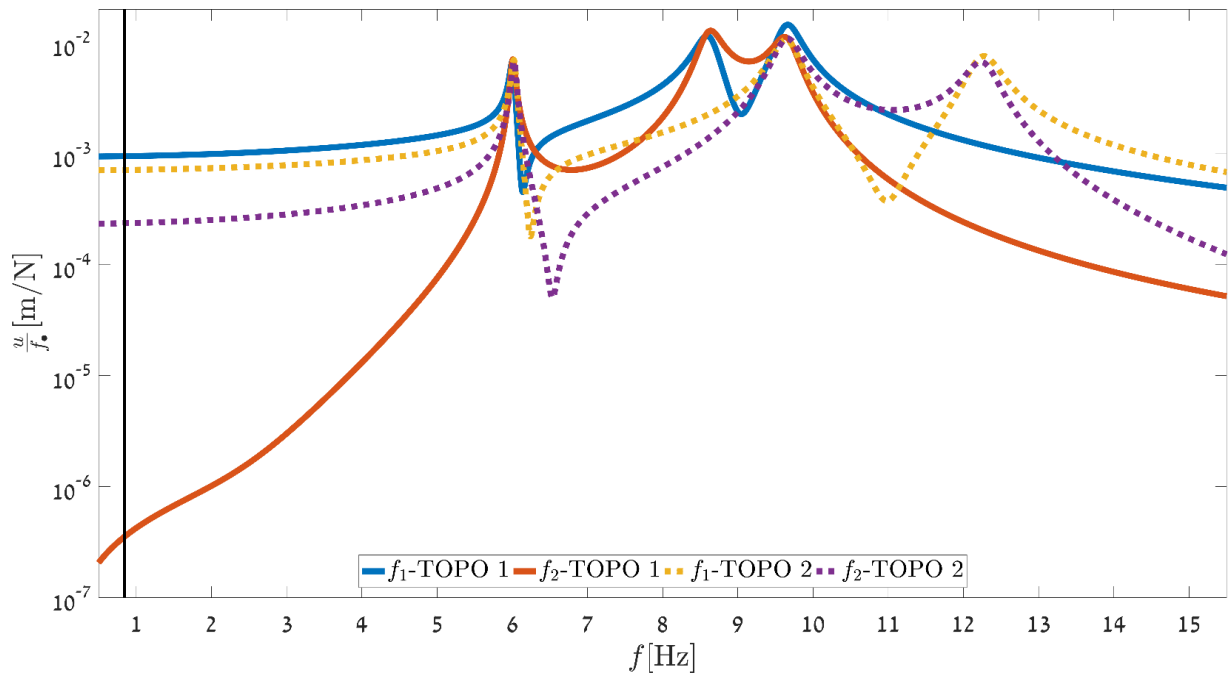

Figure 17. Linear frequency responses of the system, where the output is the $u$ coordinate and the input is either $f_1$ or $f_2$. Continuous lines / dotted lines show first topology / second topology.

## Acknowledgments

The first author would like to acknowledge the generous financial support of the Israeli Ministry of Science, Technology & Space for the Applied Science scholarship for PhD engineering students 2016.

This research sis not receive any specific grant from funding agencies in the public, commercial, or not-for-profit sectors.

# Appendix A - Additional mathematical terms

For brevity, the following mathematical terms were left outside the manuscript.

$$\boldsymbol{\kappa} = \frac{\kappa}{\tilde{\omega}^2} \begin{pmatrix} \Phi_{21}^4 & \Phi_{21}^3 \Phi_{22} & \Phi_{21}^3 \Phi_{23} \\ 3\Phi_{21}^3 \Phi_{22} & 3\Phi_{21}^2 \Phi_{22}^2 & 3\Phi_{21}^2 \Phi_{22} \Phi_{23} \\ 3\Phi_{21}^2 \Phi_{22}^2 & 3\Phi_{21} \Phi_{22}^3 & 3\Phi_{21} \Phi_{22}^2 \Phi_{23} \\ \Phi_{21} \Phi_{22}^3 & \Phi_{22}^4 & \Phi_{22}^3 \Phi_{23} \\ 3\Phi_{21}^3 \Phi_{23} & 3\Phi_{21}^2 \Phi_{22} \Phi_{23} & 3\Phi_{21}^2 \Phi_{23}^2 \\ 6\Phi_{21}^2 \Phi_{22} \Phi_{23} & 6\Phi_{21} \Phi_{22}^2 \Phi_{23} & 6\Phi_{21} \Phi_{22} \Phi_{23}^2 \\ 3\Phi_{21} \Phi_{22}^2 \Phi_{23} & 3\Phi_{22}^3 \Phi_{23} & 3\Phi_{22}^2 \Phi_{23}^2 \\ 3\Phi_{21}^2 \Phi_{23}^2 & 3\Phi_{21} \Phi_{22} \Phi_{23}^2 & 3\Phi_{21} \Phi_{23}^3 \\ 3\Phi_{21} \Phi_{22} \Phi_{23}^2 & 3\Phi_{22}^2 \Phi_{23}^2 & 3\Phi_{22} \Phi_{23}^3 \\ \Phi_{21} \Phi_{23}^3 & \Phi_{22} \Phi_{23}^3 & \Phi_{23}^4 \end{pmatrix}^T \quad \tilde{\boldsymbol{\eta}} = \begin{Bmatrix} \eta_1^3 \\ \eta_1^2 \eta_2 \\ \eta_1 \eta_2^2 \\ \eta_2^3 \\ \eta_1^2 \eta_3 \\ \eta_1 \eta_2 \eta_3 \\ \eta_2^2 \eta_3 \\ \eta_1 \eta_3^2 \\ \eta_2 \eta_3^2 \\ \eta_3^3 \end{Bmatrix} \quad (A.1)$$

$$D_0^2 \eta_{11} + \chi_1^2 \eta_{11} = - \left( \begin{pmatrix} 2i\zeta_1 \chi_1 \chi_r \left( \Lambda_{11} e^{i\varphi_1} + \Lambda_{12} e^{i\varphi_2} \right) + \\ \dfrac{3\kappa \Phi_{21} H_1}{\tilde{\omega}^2} \left( 2 A_1 \bar{A}_1 \Phi_{21}^2 + H_1^2 + 2 H_2^2 \right) \left( e^{i\varphi_1} + e^{i\varphi_2} \right) \end{pmatrix} e^{i\chi_r \tau} + \\ \dfrac{\Phi_{21}^2 k_1 \gamma_b A_1}{2 \tilde{\omega}^2} e^{i((\chi_1 - \chi_b)\tau - \varphi_b)} \right) + \text{AOT} + \text{CC},$$

$$D_0^2 \eta_{21} + \chi_2^2 \eta_{21} = - \begin{pmatrix} \begin{pmatrix} 2i\zeta_2 \chi_2 \chi_r \left( \Lambda_{21} e^{i\varphi_1} + \Lambda_{22} e^{i\varphi_2} \right) + \\ \dfrac{3\kappa \Phi_{22} H_1}{\tilde{\omega}^2} \left( 2 A_1 \bar{A}_1 \Phi_{21}^2 + H_1^2 + 2 H_2^2 \right) \left( e^{i\varphi_1} + e^{i\varphi_2} \right) \end{pmatrix} e^{i\chi_r \tau} + \\ \dfrac{\Phi_{21} \Phi_{22} k_1 \gamma_b A_1}{2 \tilde{\omega}^2} e^{i((\chi_1 - \chi_b)\tau - \varphi_b)} + \\ \dfrac{\Phi_{22} k_1}{2 \tilde{\omega}^2} \left( \left( H_1 e^{i\varphi_1} + H_2 e^{i\varphi_2} \right) \gamma_b e^{i((\chi_b + \chi_r)\tau + \varphi_b)} + \gamma_a \Phi_{21} \bar{A}_1 e^{i((\chi_a - \chi_1)\tau + \varphi_a)} \right) + \\ \dfrac{3\Phi_{21} \Phi_{22} \kappa A_1}{\tilde{\omega}^2} \left( A_1 \bar{A}_1 \Phi_{21}^2 + 2 H_1^2 + 2 H_2^2 \right) e^{i\chi_1 \tau} \end{pmatrix} + \text{AOT} + \text{CC},$$

$$D_0^2 \eta_{31} + \chi_3^2 \eta_{31} = - \begin{pmatrix} \begin{pmatrix} 2i\zeta_3 \chi_3 \chi_r \left( \Lambda_{31} e^{i\varphi_1} + \Lambda_{32} e^{i\varphi_2} \right) + \\ \dfrac{3\kappa \Phi_{23} H_1}{\tilde{\omega}^2} \left( 2 A_1 \bar{A}_1 \Phi_{21}^2 + H_1^2 + 2 H_2^2 \right) \left( e^{i\varphi_1} + e^{i\varphi_2} \right) \end{pmatrix} e^{i\chi_r \tau} + \\ \dfrac{\Phi_{21} \Phi_{23} k_1 \gamma_b A_1}{2 \tilde{\omega}^2} e^{i((\chi_1 - \chi_b)\tau - \varphi_b)} + \\ \dfrac{\Phi_{23} k_1}{2 \tilde{\omega}^2} \left( \left( H_1 e^{i\varphi_1} + H_2 e^{i\varphi_2} \right) \gamma_b e^{i((\chi_b + \chi_r)\tau + \varphi_b)} + \gamma_a \Phi_{21} \bar{A}_1 e^{i((\chi_a - \chi_1)\tau + \varphi_a)} \right) + \\ \dfrac{3\Phi_{21} \Phi_{23} \kappa A_1}{\tilde{\omega}^2} \left( A_1 \bar{A}_1 \Phi_{21}^2 + 2 H_1^2 + 2 H_2^2 \right) e^{i\chi_1 \tau} \end{pmatrix} + \text{AOT} + \text{CC}. \quad (A.2)$$

Here, AOT stands for all other terms.

# Appendix B - Modal update via nonlinear optimization problem

The reconstructed matrices topology is forced to match the model by solving two unrelated optimization problems [21]. First, the mass matrix is reconstructed by solving the following problem:

$$\mathbf{M} = m_1 \begin{pmatrix} 1 & 1 & 0 \\ 1 & 1 & 0 \\ 0 & 0 & 0 \end{pmatrix} + m_2 \begin{pmatrix} 1 & 0 & -1 \\ 0 & 0 & 0 \\ -1 & 0 & 1 \end{pmatrix} + m_3 \begin{pmatrix} 1 & 0 & 0 \\ 0 & 0 & 0 \\ 0 & 0 & 0 \end{pmatrix} = \sum_{i=1}^{3} m_n W_{m,i}$$

$$J_m = \left\| \Phi^{-T} \Phi^{-1} - \sum_{i=1}^{3} m_i W_{m,i} \right\|_F.$$

(B.1)

Where, $m_\bullet$ are solved by minimizing $J_m$, and $\|\ \|_F$ represents the Frobenius norm. In an analogous manner, the stiffness matrix is reconstructed by solving:

$$\mathbf{K} = k_1 \begin{pmatrix} 0 & 0 & 0 \\ 0 & 1 & 0 \\ 0 & 0 & 0 \end{pmatrix} + k_2 \begin{pmatrix} 0 & 0 & 0 \\ 0 & 0 & 0 \\ 0 & 0 & 1 \end{pmatrix} + k_3 \begin{pmatrix} 1 & 0 & 0 \\ 0 & 0 & 0 \\ 0 & 0 & 0 \end{pmatrix} + k_4 \begin{pmatrix} 0 & 0 & 0 \\ 0 & 1 & 1 \\ 0 & 1 & 1 \end{pmatrix} = \sum_{i=1}^{4} k_i W_{k,i}$$

$$J_k = \left\| \Phi^{-T} \Omega^2 \Phi^{-1} - \sum_{i=1}^{4} k_i W_{k,i} \right\|_F.$$

(B.2)

Where, $k_\bullet$ are solved by minimizing $J_k$.

Although the solution of Eq.(B.1) and Eq.(B.2) forces the topology, the resulting natural frequencies and normal modes differ significantly from the ones estimated by the SDT, see Table 2. To overcome the problem, a nonlinear optimization problems is suggested, which couples $m_\bullet$ and $k_\bullet$ to minimize the natural frequencies and normal modes estimation error.

In the following problem, the parameters are $m_\bullet$ and $k_\bullet$:

$$\min_{m_\bullet, k_\bullet} \quad J_{mk} = \left\| \hat{\boldsymbol{\omega}} - \boldsymbol{\omega} \right\|_2 + \left\| \hat{\boldsymbol{\Phi}} - \boldsymbol{\Phi} \right\|_2 \quad \text{s.t.} \quad \begin{cases} 0.5 k_{\bullet 0} \leq k_\bullet \leq 1.5 k_{\bullet 0} \\ 0.5 m_{\bullet 0} \leq m_\bullet \leq 1.5 m_{\bullet 0} \\ \left( \mathbf{K} - \hat{\omega}_\bullet^2 \mathbf{M} \right) \hat{\phi}_\bullet = 0 \end{cases}.$$

(B.3)

Where $\hat{\boldsymbol{\omega}}$ and $\hat{\boldsymbol{\Phi}}$ are the natural frequencies vector and modal matrix of the updated system, $\boldsymbol{\omega}$ and $\boldsymbol{\Phi}$ are the estimated natural frequencies and modal matrix computed by the SDT and $m_{\bullet 0}$ and $k_{\bullet 0}$ are the solution of Eq.(B.1) and Eq.(B.2).

The various problems gradients are computed as follows:

$$\frac{\partial J_{mk}}{\partial p} = \frac{\sum_{n=1}^{3} (\hat{\omega}_n - \omega_n)(\partial \hat{\omega}_n / \partial p)}{\sqrt{\sum_{n=1}^{3} (\hat{\omega}_n - \omega_n)^2}} + \frac{\sum_{n=1}^{9} (\hat{\phi}_n - \phi_n)(\partial \hat{\phi}_n / \partial p)}{\sqrt{\sum_{n=1}^{9} (\hat{\phi}_n - \phi_n)^2}},$$

(B.4)

where $p$ can be each parameter and $\partial \hat{\omega}_n / \partial p$ and $\partial \hat{\phi}_n / \partial p$ are provided in [22].

The problem is nonlinear and has multiple minimizers which can be found with various solvers. Out of the possible minimizers, the one closest to the solution of the linear problems is chosen, and the results for both topologies are summarized in Table 2.

| Topology | Problem | $\hat{\boldsymbol{\omega}}$ Hz | $\boldsymbol{\omega}$ Hz | $\|\hat{\boldsymbol{\omega}} - \boldsymbol{\omega}\|$ Hz |
|---|---|---|---|---|
| 1 | Nonlinear | 6.0081,  8.6203,  9.6518 | 6.0081,  8.6203,  9.6518 | $O(10^{-8})$,  $O(10^{-8})$,  $O(10^{-8})$ |
| 1 | Linear | 5.7577,  7.6596,  8.9585 | | $O(10^{-1})$,  $O(10^{-1})$,  $O(10^{-1})$ |
| 2 | Nonlinear | 6.0252,  9.6524,  12.2598 | 6.0252,  9.6524,  12.2598 | $O(10^{-8})$,  $O(10^{-8})$,  $O(10^{-8})$ |
| 2 | Linear | 5.8126,  9.1252,  10.7194 | | $O(10^{-1})$,  $O(10^{-1})$,  $O(1)$ |

| Topology | Problem | $\hat{\boldsymbol{\Phi}}$ kg-1/2 | $\boldsymbol{\Phi}$ kg-1/2 | $\|\hat{\boldsymbol{\phi}} - \boldsymbol{\phi}\|$ kg-1/2 |
|---|---|---|---|---|
| 1 | Nonlinear | $\begin{bmatrix} 0.3456 & 0.0335 & 0.2377 \\ 0.3076 & 1.0619 & -1.3417 \\ -0.3440 & 1.2785 & 1.0650 \end{bmatrix}$ | $\begin{bmatrix} 0.3383 & 0.0364 & 0.2701 \\ 0.2744 & 1.1143 & -1.3568 \\ -0.2999 & 1.3228 & 1.0823 \end{bmatrix}$ | $O(10^{-2})$,  $O(10^{-2})$,  $O(10^{-2})$ |
| 1 | Linear | $\begin{bmatrix} 0.3105 & 0.0369 & 0.2827 \\ 0.3720 & 1.0020 & -1.1679 \\ -0.4368 & 1.1084 & 0.9640 \end{bmatrix}$ | | $O(10^{-1})$,  $O(10^{-1})$,  $O(10^{-1})$ |
| 2 | Nonlinear | $\begin{bmatrix} 0.3463 & 0.2383 & 0.0057 \\ 0.3209 & -1.1787 & -1.2386 \\ -0.3303 & 1.2487 & -1.1311 \end{bmatrix}$ | $\begin{bmatrix} 0.3389 & 0.2718 & 0.0031 \\ 0.2840 & -1.1826 & -12740 \\ -0.2932 & 1.12674 & -1.1621 \end{bmatrix}$ | $O(10^{-2})$,  $O(10^{-2})$,  $O(10^{-2})$ |
| 2 | Linear | $\begin{bmatrix} 0.3180 & 0.2766 & 0.0210 \\ 0.3674 & -0.9754 & -1.1956 \\ -0.3938 & 1.1669 & -0.9241 \end{bmatrix}$ | | $O(10^{-1})$,  $O(10^{-1})$,  $O(10^{-1})$ |

Table 2. Estimated parameters.